\documentclass[letterpaper,aps,prd,preprint,showpacs,nofootinbib,superscriptaddress]{revtex4}
\usepackage{ulem}
\usepackage{graphicx}
\usepackage{amsmath,amssymb,mathrsfs}
\usepackage{bbm}
\usepackage{color}
\usepackage{xcolor}
\usepackage{ulem}
\usepackage{dsfont}
\usepackage{cancel}
\usepackage{dsfont}
\usepackage{epstopdf}
\usepackage{epsfig}
\usepackage{bm}
\usepackage{multirow}
\usepackage{hyperref}
\usepackage{enumitem}
\usepackage{footnote}

\newcommand{\beq}{\begin{equation}}
\newcommand{\eeq}{\end{equation}}
\newcommand{\bea}{\begin{eqnarray}}
\newcommand{\eea}{\end{eqnarray}}

\definecolor{Red}{rgb}{1.,0.,0.}

\begin{document}
\vspace*{2cm}
\title{Phenomenology of  2HDM with VLQs}

\author{A. Arhrib}
\email[]{aarhrib@gmail.com}
\affiliation{\small Facult\'e des Sciences et Techniques, Abdelmalek Essaadi University, B.P. 416, Tangier, Morocco}
\affiliation{\small Physics Division, National Center for Theoretical Sciences, Hsinchu, Taiwan}
\author{R. Benbrik}
\email[]{r.benbrik@uca.ac.ma}
\affiliation{\small MSISM Team, Facult\'e Polydisciplinaire de Safi,
Sidi Bouzid, BP 4162,  Safi, Morocco}
\author{S.J.D. King}
\email[]{sjd.king@soton.ac.uk}
\affiliation{\small School of Physics and Astronomy, University of Southampton,
  Southampton, SO17 1BJ, United Kingdom}

\author{B. Manaut}
\email[]{b.manaut@usms.ma}
\affiliation{\small D\'epartement de Physique, LIRST, Facult\'e Polydisciplinaire, Universit\'e Sultan Moulay Slimane, B\'eni Mellal, Morocco}
\author{S. Moretti}
\email[]{stefano@soton.ac.uk}
\affiliation{\small School of Physics and Astronomy, University of Southampton,
Southampton, SO17 1BJ, United Kingdom}

\author{C.S. Un}
\email[]{csalih@zewailcity.edu.eg}
\affiliation{\small Center for Fundamental Physics, Zewail City of Science and Technology, 6 October City, Giza, Egypt}
\affiliation{\small Department of Physics, Uluda\~{g} University, 16059 Bursa, T\"{u}rkiye}

%\date{\today}
\vspace*{-3.5cm}
\begin{abstract}
\footnotesize
In this paper, we examine the consistency of the Large Hadron Collider (LHC) 
 data collected during Run 1 and 2  by the ATLAS and CMS experiments with the predictions 
of  a 2-Higgs Doublet Model (2HDM) embedding Vector-Like Quarks (VLQs) 
for $pp \to H,A$ production and $H,A\to\gamma\gamma$ decay mechanisms, respectively, of (nearly)
degenerate CP-even ($H$) and CP-odd ($A$) Higgs bosons.  
We show that a scenario
containing one single VLQ with Electro-Magnetic (EM) charge $2/3$ can 
explain the above ATLAS and CMS data for masses in
the region 350 GeV $\leq  m_{\rm VLQ}\leq 1.5$ TeV or so, depending on 
$\tan\beta$, and for several values of the mixing angle between the
top quark ($t$) and its VLQ counterpart ($T$).
We then perform a global fit onto the model by including all relevant experimental as well as theoretical constraints. 
The surviving samples of our analysis are discussed within 2$\sigma$ of the LHC measurements.
Additionally, we also comment on the recent anomalous result reported by CMS using Run
2 data on the associated Standard Model (SM) Higgs boson production with top quark pairs $pp\to t\bar th$
 with an observed significance of 3.3$\sigma$. Other than these
specific examples, we also present a phenomenological analysis of the main features of the model, including 
the most promising  $T$ decay channels.

\end{abstract}
\vspace*{-1.5cm}

\maketitle
%%%%%%%%%%%%%%%%%%%%%%%%%%%%%%%%%%%%%%%%
%%%%%%%%%%%%%%%%%%%%%%%%%%%%%%%%%%%%%%%%
%{Introduction}
\section{Introduction}
After the Higgs boson discovery in  Run 1 of the Large Hadron Collider (LHC) at CERN \cite{Aad:2012tfa,Chatrchyan:2012xdj}, the ATLAS and CMS collaborations have carried out a broad programme looking for  new physics Beyond the Standard Model (BSM) at the TeV scale. In particular, they have developed a powerful detection machinery of spin-0 resonances. However, new physics phenomena might take different forms from those established so far and, once discovered, they will require a complementary effort in order to understand their underlying nature. In fact, there are already several potential anomalies in the Run 1 Higgs data indicating possible deviations from the SM expectations in the Higgs sector. In particular, the signal strength of the $t\bar{t}h$ associated production mode is the most prominent one, while milder effects are seen in the fits to  those extracted from the other production modes active 
at the LHC, i.e., gluon-gluon scattering (ggh), Higgs-strahlung (hV) and Vector Boson Fusion (VBF).

A possibility to capture at once all such anomalies is offered by the presence of Vector-Like Quark (VLQs) as they can, on the one hand, affect the SM-like Higgs (henceforth denoted by $h$) production and decay phenomenology (as they would enter the loops mediating the processes $gg\to h$ as well as $h\to \gamma\gamma$ and $Z\gamma$) and, on the other hand, mediate a 
$t\bar t h$ final state (tth). Intriguingly, the same VLQs affecting $h$ processes would also do so for heavier Higgs states, which may pertain to a BSM scenario. To be specific in defining our framework, we investigate the effects of VLQs in the production and decay of  Higgs bosons within a 2-Higgs Doublet Model type-II (2HDM-II). In practice, we concentrate here on new  states of matter that are
heavy spin 1/2 particles that transform as triplets under color but,
unlike SM quarks, their left- and right-handed couplings have the same
Electro-Weak (EW) quantum numbers. Furthermore, their couplings to Higgs bosons
do not participate in the standard EW Symmetry Breaking (EWSB)
dynamics onset by the Higgs mechanism, hence they are not of
Yukawa type (i.e., proportional to the mass), rather they are
additional parameters, which can then be set as needed in order to achieve both compliance with present data and predict testable signals for the future.

The ATLAS and CMS collaborations, while collecting data at 7, 8 and 13 TeV, performed searches for  
VLQs with different quantum numbers, probing single and pair production
mechanisms, as well as decay modes into all three generations of SM quarks
(for the most updated experimental results of ATLAS and CMS we refer to the
respective web pages \cite{twikiATLAS8TeV,twikiATLAS13TeV,twikiCMS}). However,
new extra quarks can be charged under new symmetries, like $T$-parity in
Little Higgs \cite{ArkaniHamed:2002qx,Cheng:2003ju,Cheng:2004yc,
Low:2004xc,Hubisz:2004ft,Cheng:2005as,Hubisz:2005tx} and Kaluza-Klein
parity in Extra Dimension 
\cite{Antoniadis:1990ew,Appelquist:2000nn,Servant:2002aq,
Csaki:2003sh,Cacciapaglia:2009pa} models. 
%\textcolor{red}{
Such VLQs have been searched for at both the
Tevatron \cite{Aaltonen:2011rr, Aaltonen:2011na} and LHC
\cite{ATLAS:2011mda,CMS:2012dwa}, though no evidence for the existence of other quarks, beside those of the SM, has been obtained. Direct bounds on heavy chiral quarks can be interpreted as bound on VLQs, but it must be stressed that decay channels of VLQs are different from decay channels of heavy chiral quarks \cite{Barducci:2017xtw}. Thus, if the VLQs have a strong mass degeneracy,
the visible decay products of the VLQ are too soft to be detected and, as a
consequence, the bounds on the VLQ mass can be very weak, analogous to the
case of strong degeneracy between squarks and neutralinos in
supersymmetry. Intriguingly, as we shall detail below, even in our simple 
scenario, VLQ mass values down to 400 GeV are still possible,
so that they could strongly affect, e.g., the $gg\to h \to V\gamma $ (with $V = \gamma, Z$)
rates.
%}

{Now, let us also assume that additional
(pseudo)scalar objects possibly behind the LHC experimental data do
originate from the same EWSB mechanism governing the generation of the
$ \approx 125$ GeV Higgs state. This is indeed a possibility not excluded by
current theoretical and experimental constraints. 
Under these circumstances, it is then of
phenomenological importance to consider the case of a second Higgs doublet
participating in EWSB alongside the one responsible for the discovered Higgs
state. This mass generation dynamics is well known in the form of 2HDMs \cite{Branco:2011iw}. 
We are therefore left with a new
physics construct that would include a 2HDM supplemented by one or more VLQs
as a potential scenario that could accommodate the LHC data on the $\approx 125$ GeV Higgs boson and additionally explain results above the SM yield.

In this paper, we wish to build on the results of
\cite{Benbrik:2015fyz,Aguilar-Saavedra:2013qpa,Badziak:2015zez,
  Angelescu:2015uiz}, where a similar possibility was discussed (by some of
us), in which the role of a 2HDM was played by a SM-like Higgs doublet supplemented by  
an additional Higgs singlet. We intend to review here a 2HDM plus single VLQ
scenario, where the VLQ has the same Electro-Magnetic (EM) charge of the top quark (with which
it then mixes), as a candidate to study the implications of the heavy Higgs searches by the ATLAS and CMS collaborations
both with 8 TeV and the latest 13 TeV data. Furthermore, we will relate such data samples to those
involving  $ \gamma\gamma $ and $ Z\gamma $ final states. In addition, we will show an enhancement of the $pp\to t\bar t h$ cross-section at the LHC induced by small mixing of the top quark with the additional state $T$.
Finally, we will discuss the possibility of VLQs  produced as real objects in the detector 
decaying into Higgs boson states, both neutral and charged.

Our paper is formatted as follows. In the
next section, we describe in some detail the model concerned. In the three
subsequent sections, we present our results, followed by our conclusions.
}
%%%%%%%%%%%%%%%%%%%%%%%%

%%%%%%%%%%%%%%%%%%%%%%%%%%%%%%%%%%%%%%%%
\section{A 2HDM extended by an up-type vector-like quark}
\label{sec:II}
%
%%%%%%%%%%%%%%%%%%%%%%%%%%%%%%%%%%%%%%%%
%
A simple extension of the SM is the well-known 2HDM
that expands the Higgs sector of the SM by an additional Higgs doublet.
The spectrum of the model contains additional Higgses and possesses an alignment
limit \cite{Carena:2013ooa}, in which one of the Higgses
completely mimics the SM one. \\
To describe our model, we start with the well known CP-conserving
 2HDM scalar potential for ($\Phi_1, \Phi_2)$ with a discrete symmetry
$\Phi_1\to -\Phi_1$ that is only violated softly by dimension
two terms \cite{Branco:2011iw,Gunion:1989we}:
\begin{eqnarray}
V\left( \Phi_1, \Phi_2 \right)  &=& m_{11}^2 \Phi_1^\dag \Phi_1 + m_{22}^2 \Phi_2^\dag \Phi_2 - m_{12}^2 \left( \Phi_1^\dag \Phi_2 + \Phi_2^\dag \Phi_1 \right) \nonumber \\
&+& \frac{\lambda_1}{2} \left( \Phi_1^\dag \Phi_1 \right)^2 + \frac{\lambda_2}{2} \left( \Phi_2^\dag \Phi_2 \right)^2
+ \lambda_3 \left( \Phi_1^\dag \Phi_1 \right) \left( \Phi_2^\dag \Phi_2 \right)  \nonumber\\
&+& \lambda_4 \left( \Phi_1^\dag \Phi_2 \right) \left( \Phi_2^\dag \Phi_1 \right) + \frac{\lambda_5}{2} \left[ \left( \Phi_1^\dag \Phi_2 \right)^2 + \left( \Phi_2^\dag \Phi_1 \right)^2 \right],
\label{thdmV}
\end{eqnarray}
where all parameters are real.
The two complex scalar doublets $\Phi_{1,2}$ may be rotated into a
basis, $H_{1,2}$, where only one obtains a Vacuum Expectation Value (VEV),
\begin{eqnarray}
 H_1=\left(
   \begin{array}{c}
     G^+ \\
     \frac{v+\varphi^0_1+iG^0}{\sqrt{2}} \\
   \end{array}
 \right),\quad  H_2=\left(
   \begin{array}{c}
     H^+ \\
     \frac{\varphi^0_2+iA}{\sqrt{2}} \\
   \end{array}
 \right),
\end{eqnarray}
where $G^0$ and $G^\pm$ are the would-be Goldstone bosons and $H^\pm$ are a pair of
charged Higgses. Herein, $A$ is a CP-odd pseudoscalar which does not mix with
the other neutral states in the CP-conserving case.
The physical CP-even scalars $h$ and $H$ are
 mixtures of $\varphi^0_{1,2}$ and the scalar mixing is parameterized as\footnote{Hereafter, $s_X\equiv\sin X$ and $c_X=\cos X$.}
\begin{eqnarray}
\left(
 \begin{array}{c}
 h\\
         H \\
   \end{array}
 \right)= \left(
          \begin{array}{cc}
            s_{\beta-\alpha} & c_{\beta-\alpha} \\
            c_{\beta-\alpha} & -s_{\beta-\alpha} \\
          \end{array}
        \right)\left(
   \begin{array}{c}
 \varphi^0_1\\
 \varphi^0_2 \\
   \end{array}
 \right),
\end{eqnarray}
where $\tan\beta=v_2/v_1$ is the angle used to rotate $\Phi_{1,2}$ into
$H_{1,2}$ and $\alpha$ is the additional mixing needed to diagonalize
 the CP-even mass matrix.
As mentioned in the introduction,
in order to alter the gluon-gluon-Higgs, photon-photon-Higgs and/or $Z$-photon-Higgs
 couplings, one can advocate the inclusion of new heavy fermions such as a VLQ partner of the top quark
with the same EM charge.
In fact, there are many SM extensions that require vector-like fermions
in their spectrum (for an overview see
\cite{Ellis:2014dza,Aguilar-Saavedra:2013qpa}).
Such a new VLQ will mix with the top quark through the Yukawa interactions
 and can contribute, therefore, to some  SM observables.
%Although the introduced new quark it can couple to the
%SM particles through the mixings from the Yukawa sector,
%scalar potential, and gauge sector.
To derive these new interactions, we first study the Yukawa sector
within a 2HDM extended by a VLQ pair $(T_{L},T_{R})$
in the $\mathbf{1}_{2/3}$ representation of the SM EW group.
%Models presenting extra heavy quarks find a justification as
%\blue{*collider at LHC machine*?}.
In the 2HDM-II, our concern here,
one doublet couples to up quarks and the other one couples to down quarks and charged leptons. The most general
renormalizable model for the quark Yukawa interactions and mass terms
can be described, limited to third generation quarks and new VLQs,  by the following Lagrangian,
 \begin{eqnarray}
-\mathcal{L}_Y^{II} &\supset& y_T\overline{Q_L^0} \widetilde{H}_2T_R^0+
\xi_T\overline{Q^0_L} \widetilde{H}_1T_R^0+M_T \overline{T_L^0}T_R^0 \nonumber\\
&\supset& y_T ( \overline{t_L^0}, \overline{b_L^0})\left(
   \begin{array}{c}
      \frac{\varphi^0_2-iA}{\sqrt{2}} \\
                  -H^- \\
         \end{array}
    \right)T_R^0 +\xi_T ( \overline{t_L^0}, \overline{b_L^0})\left(
            \begin{array}{c}
       \frac{v+\varphi^0_1-iG^0}{\sqrt{2}} \\
            -G^- \\
      \end{array}
   \right)T_R^0+M_T\overline{T_L^0}T_R^0 + \rm{h.c},
\end{eqnarray}
where $\tilde H_i \equiv i \tau_2 H^*_i$ ($i=1,2$),
$Q^0_{L}$ are the SM quark doublets and the $u^i_{R}$'s are the 
SM up-type quark singlets.
Note that additional kinetic mixing terms of the form
$\overline{T_{L}}t^i_{R}$
 can always be rotated away and reabsorbed into the definition of $y_{t,T}$.
Furthermore, one can, without loss of generality, choose a weak interaction
basis where $y_t$ is diagonal and real.  In the weak eigenstate basis
$(t_L^0,T_L^0)$,  the top quark and VLQ mass matrix is
\begin{eqnarray}
{\mathcal M}=\left(
\begin{array}{cc}
    \frac{y_t v}{\sqrt{2}} & \frac{\xi_T v}{\sqrt{2}} \\
    0 & M_T \\
  \end{array}
        \right),
\label{eq:massmat}
\end{eqnarray}
where $y_t$ and $\xi_T$ are the Yukawa couplings for the top quark and VLQ,
respectively, $v=246$ GeV is the VEV of the SM Higgs doublet while
$M_T$ is a bare mass term of the VLQ, which, as intimated, is 
unrelated to the Higgs mechanism of EWSB.
It is clear from the above mass matrix that the physical mass of
the heavy top, $m_T$, is different from $M_T$ due to the $t-T$ mixing.
Furthermore, such a mass matrix can be diagonalized by a 
bi-unitary transformation such that
\begin{eqnarray}
U_L\mathcal{M}U_R^\dag=\mathcal{M}_{d},\quad\quad 
U_{R,L}\mathcal{M}^{\dag}\mathcal{M}U_{R,L}^\dag=\mathcal{M}_{d}^2,
\label{equa:diag}
\end{eqnarray}
with $\mathcal{M}$ the matrix given in Eq. (\ref{eq:massmat}) and
$\mathcal{M}_d$ the diagonalized one. The unitary matrices
$U_L$ and $U_R$ are defined by
\begin{eqnarray}
\left(
\begin{array}{c}
 t_{L,R} \\
 T_{L,R}  \\
    \end{array}
\right)=\left( \begin{array}{cc}
\cos(\theta_{L,R}) &  -\sin(\theta_{L,R}) \\
\sin(\theta_{L,R}) &  \cos(\theta_{L,R}) \\
\end{array}
\right)\left(
\begin{array}{c}
t_{L,R}^0 \\
T_{L,R}^0  \\
\end{array}
\right) .
\end{eqnarray}

In fact, the mixing angles $\theta_L$ and $\theta_R$ are not
independent parameters. From the bi-unitary
transformations applied to Eq. (\ref{equa:diag}), we can derive
the following relations:
\begin{eqnarray}
\tan(2\theta_L)=\frac{4m_tM_T}{2M_T^2-2m_t^2-\xi_T^2v^2}, \quad\quad
\tan(2\theta_R)=\frac{2\sqrt{2}\xi_T m_tv}{2M_T^2+2m_t^2-\xi_T^2v^2} .
  \end{eqnarray}
%The Yukawa sector is determined in terms of the couplings
%$y_t$, $y_T$ and $\xi_T$ using Tr$(U_L \mathcal{M}\mathcal{M}^\dag U_L^\dag)$,
%Det$(U_L \mathcal{M}\mathcal{M}^\dag U_L^\dag)$ as well as the condition
%on off-diagonal elements of $U_L \mathcal{M}\mathcal{M}^\dag U_L^\dag$
%to be zero.
The above equations in turn give the following relationships between
$\theta_L$ and $\theta_R$, see \cite{Aguilar-Saavedra:2013qpa}:
\begin{eqnarray}
\tan\theta_L = \frac{m_t}{m_T} \tan\theta_R, \quad\quad
\quad\quad \frac{\xi_T}{y_t} = s_L c_L \frac{m^2_T - m^2_t}{m_t m_T}.
\end{eqnarray}
%The new quark, $T$, effects provide better constraints
%only on the mixing of the vector-like singlet quark with the top quark.
%
%Neglecting the small mixing with the first two generations the
%$T-t$ interaction can be described by three independent physical
%parameters: two quark masses : $m_t, m_T$ and a
%mixing angle ($\sin\theta_{L}$).

After rotating the weak eigenstates $(t_L^0,T_L^0)$ into the mass eigenstates,
the Yukawa Lagrangian takes the following form:
\begin{eqnarray}
-\mathcal{L}_Y^{II} &\supset& (\overline{t_L},\overline{T_L})U_L
\left[\varphi^0_1\left(
 \begin{array}{cc}
 \frac{y_t}{\sqrt{2}} & \frac{\xi_T}{\sqrt{2}} \\
  0 & 0 \\
 \end{array}
 \right)+\varphi^0_2\left(
\begin{array}{cc}
  \frac{y_t}{\sqrt{2}} \xi_u & \frac{y_T}{\sqrt{2}} \\
         0 & 0 \\
 \end{array}
 \right)\right]U_R^\dag \left(
 \begin{array}{c}
    t_R \\  T_R  \\
 \end{array}
 \right)\nonumber\\
   &-& i (\overline{t_L},\overline{T_L})A \left[U_L\left(
  \begin{array}{cc}
  \frac{y_t}{\sqrt{2}} \xi_u & \frac{y_T}{\sqrt{2}} \\
                     0 & 0 \\
   \end{array}
  \right)U_R^\dag\right]\left(
  \begin{array}{c}  t_R \\  T_R \\
 \end{array}
 \right) + \rm {h.c}.
\label{equa:intr}
\end{eqnarray}
The neutral Higgs couplings to top ($t$) and heavy top ($T$) 
quark pairs normalized to the $h_{\rm SM}t\bar{t}$ one are given in Appendix A.

In our 2HDM+VLQ scenario, neutral and charged current interactions receive contributions from the new VLQ,
\begin{eqnarray}
\mathcal{L} &=& -\frac{g}{\cos\theta_W} Z_\mu \bar{f} \gamma^\mu \left(g^L_{ff'} P_L + g^R_{ff'} P_R \right)f' + \frac{g}{\sqrt{2}}\left(V_{tb} \bar{t} + V_{Tb} \bar{T}\right) \gamma^\mu P_L b W^{+}_{\mu} + {\rm h.c},
\end{eqnarray}
with $f,f' = t, T$. The new couplings are modified as follows:
\begin{eqnarray}
g^L_{tt} &=& T^t_3 - Q_t  \sin^2\theta_W - \frac{s^2_L}{2}, \\
g^L_{TT} &=& T^T_3 - Q_T  \sin^2\theta_W - \frac{c^2_L}{2}, \\
g^L_{tT} &=& \frac{s_L c_L}{2}, \quad g^R_{tt} = g^R_{TT} = -Q_t \sin^2\theta_W, \quad g^R_{tT} = 0,\\
V_{Tb}  &=& s_L\, \, , {\rm and } \, \, V_{tb} = c_L .
\end{eqnarray}
Finally, the interaction with the charged salar boson and the new quark $T$ can be written as
\begin{eqnarray}
\mathcal{L} &=& -\frac{g V_{Tb}}{\sqrt{2} M_W}\bar{T}\left(\frac{m_T}{\tan\beta} P_L + m_b \tan\beta P_R \right) b H^++ {\rm h.c}.
\end{eqnarray}

%%%%%%%%%%%%%%%%%%%%%%%%%%%%%%%%%%%%%%%%%%%%
\section{Results and discussions}
In our numerical calculation, we consider the scenario with a light Higgs boson $h$ as the SM-like state, with $m_h = 125$ GeV.  We take into account theoretical constraints from vacuum stability, unitarity and perturbativity. We then enforce bounds from precision EW data (such as the oblique parameters $S, T$ and $U$) and adopt constraints on the charged Higgs boson mass from  the 2HDM-II using $b\to s \gamma$ rates, which set a limit $m_{H^\pm} > 580$ GeV \cite{Misiak:2017bgg}. 

In addition, we perform a global $\chi^2$ analysis for the signal strengths of the observed Higgs boson $h$ from the combined production modes $i =1$ (ggh+t{t}h) as well as $i=2$ (VBF+Vh) and decay modes into $ f = \gamma\gamma$, $ZZ$, $W^+W^-$, $\tau^-\tau^+$ and $b\bar{b}$ \cite{Khachatryan:2016vau}:
\begin{equation}\label{chi2}
\chi^2 = \sum_{i=1,2} \frac{\left(\hat{\mu}^f_i - \mu^f_i\right)^2}{\Delta\mu^{f\,2}_i}
\end{equation}
where the signal strength variable $ \mu^f_i$ is defined as
\begin{equation}
\mu^f_i = \frac{\sigma(i \to h ) {\rm BR}(h \to f )}{\sigma_{\rm SM}(i \to h ) {\rm BR}_{\rm SM}(h \to f )},
\end{equation}
in terms of a production cross-section, $\sigma$, and a decay Branching Ratio (BR).
%Here, $i$ stand for the production modes\footnote{In particular, $i=1$ corresponds to ggh and t{t}h and $i=2$ stands for VBF and $Vh$.} of the Higgs boson. 
The parameters with the subscript ``SM" represents the corresponding values for the SM. The experimentally obtained best-fit signal strength values which we have implemented in our analysis are given in Tab. \ref{tab:HIGGS_datarun12}. Furthermore, in Eq. (\ref{chi2}), $ \Delta\mu^{f\,2}_i $ represents the error associated with the experimental measurement. We use HiggsBounds-4.3.1 \cite{Bechtle:2013wla} to constrain the non-observation of neutral and/or charged Higgs bosons at the LHC at $95\%$CL.
%%%
\begin{center}
\begin{table}[t] \footnotesize
\begin{tabular}{|c||c|c|c|} \hline
\multirow{3}{*}{Higgs Signal strength} & \multicolumn{3}{c|}{LHC data} \\ \cline{2-4} &
~Run 1 \cite{Khachatryan:2016vau} & \multicolumn{2}{c|}{Run 2} \\ \cline{3-4} &
 & ATLAS \cite{Atlas:2016081,Atlas:2016063,Atlas:2016003} & CMS \cite{CMS:2016ixj,CMS:2016zbb,CMS:2016020}\\ \hline\hline
% \gamma\gamma
$ \hat{\mu}^{\gamma\gamma}_{1}$  & $1.10^{+0.23}_{-0.21} $ & $0.67^{+0.25}_{-0.21} $   & $0.77^{+0.25}_{-0.23} $\\ \hline
$ \hat{\mu}^{\gamma\gamma}_{2}$  & $0.8^{+.71}_{-0.71}$    & $2.25^{+0.75}_{-0.75} $   & $1.59^{+0.73}_{-0.45} $ \\ \hline
% ZZ
$ \hat{\mu}^{ZZ}_{1}$            & $1.27^{+0.28}_{-0.24} $ & $1.42^{+0.35}_{-0.31} $   & $1.20^{+0.22}_{-0.21} $ \\ \hline
$ \hat{\mu}^{ZZ}_{2}$            & $1.66^{+0.51}_{-0.44}$  & $3.8^{+2.8}_{-2.2} $   & $0.67^{+1.61}_{-0.67} $ \\ \hline
% WW
$ \hat{\mu}^{WW}_{1}$            & $1.06^{+0.21}_{-0.18} $ &  - & -\\ \hline
$ \hat{\mu}^{WW}_{2}$            & $1.27^{+0.53}_{-0.45} $ &  - & - \\ \hline
% b\bar{b}
$ \hat{\mu}^{b\bar{b}}_{1}$      & $0.64^{+0.37}_{-0.28} $ &  $3.9^{+2.8}_{-2.9} $  & $3.7^{+2.4}_{-2.5} $ \\ \hline
$ \hat{\mu}^{b\bar{b}}_{2}$      & $0.51^{+0.40}_{-0.37} $ &   &  \\ \hline
% tau tau
$ \hat{\mu}^{\tau \tau}_{1}$     & $1.05^{+0.33}_{-0.27} $ &  - & - \\ \hline
$ \hat{\mu}^{\tau \tau}_{2}$     & $1.24^{+0.58}_{-0.54} $ &  - &  - \\ \hline
\end{tabular}
\caption{
\label{tab:HIGGS_datarun12}
The Higgs signal strengths in various production and decay channels measured by ATLAS  and CMS 
%\cite{Khachatryan:2014jba}  
presented in combination at both LHC Run 1 (combined $\sqrt{s} = 7$ and 8 TeV) and Run 2 ($\sqrt{s} = 13$ TeV).
}
\end{table} 
\end{center}

%\begin{table}[tb!]
%\begin{center}
%\renewcommand*{\arraystretch}{1.1} %% Change the default height of the rows
%\begin{tabular}{|c|c|c|c|}
%\hline
%Signal strength   & Production mode   &   Run 1  &  Run 2  \\   
%\hline
%$ \mu^{\gamma\gamma}_{i}$   & ggh+tth & $1.10^{+0.23}_{-0.22} $  &  $0.70^{+0.19}_{-0.18}$ 
%\\ \cline{2-4} 
%         & VBF + Vh   & $0.1^{+1.1}_{-0.6}$ &  $3.8^{+2.8}_{-2.2}$      
%% 
%\\ \hline 
%$ \mu^{ZZ}_{i}$   & ggh+tth & $1.13^{+0.34}_{-0.31} $  &  $1.15^{+0.27}_{-0.23}$ 
%\\ \cline{2-4} 
%         & VBF + Vh   & $0.1^{+1.1}_{-0.6}$ &  $3.8^{+2.8}_{-2.2}$      
%% 
%\\ \hline
%$ \mu^{WW}_{i}$   & ggh+tth & $0.84^{+0.17}_{-0.17} $  &  $-$ 
%\\ \cline{2-4} 
%         & VBF + Vh   & $0.1^{+1.1}_{-0.6}$ &  $3.8^{+2.8}_{-2.2}$      
%% 
%\\ \hline
%$ \mu^{b\bar{b}}_{i}$   & ggh+tth & $0.64^{+0.37}_{-0.28} $  &  $-3.7^{+1.82}_{-1.89}$ 
%\\ \cline{2-4} 
%         & VBF + Vh   & $0.1^{+1.1}_{-0.6}$ &  $3.8^{+2.8}_{-2.2}$      
%% 
%\\ \hline
%$ \mu^{\tau\tau}_{i}$   & ggh+tth & $1.05^{+0.33}_{-0.27} $  &  $0.70^{+0.19}_{-0.18}$ 
%\\ \cline{2-4} 
%         & VBF + Vh   & $0.1^{+1.1}_{-0.6}$ &  $3.8^{+2.8}_{-2.2}$      
% 
%\\ \hline 
%
%\end{tabular}
%\quad 
%\end{center}
%\caption{
%\label{tab:HIGGS_datarun12}
%The Higgs signal strengths in various production and decay channels measured at both ATLAS  \cite{Khachatryan:2016vau}  and CMS \cite%{Khachatryan:2014jba,CMS:2016ixj,CMS:2016zbb}  presented in combination at both LHC Run 1 and  2.
%}
%\end{table}
\vspace{-2cm}
%%%%%%%%%%%%%%%%%%%%%%%%%%%%%%%%%%%%%%%%%%%%
\subsection{Constraints on $m_T$}
As intimated in the introduction, ATLAS and CMS have performed direct 
searches for VLQs at 7, 8 TeV,  having potential sensitivities up to 800 GeV or 
so \cite{XQCAT,Barducci:2014ila,Pheno,Aguilar-Saavedra:2013qpa}. We have already explained that
several VLQ scenarios may be conceived in order to enable $m_{\rm VLQ}$ values 
down to 350 GeV or so yet still compatible
with data. Clearly, the decay patterns of  new VLQs depend on the 
representation of these fermionic states. In our rather simple 
scenario, i.e., in the case of a singlet VLQ, if we neglect the first 
and second generation mixing, the heavy top $T$ will decay into the 
following final states:  $W^+ b, Z t$ and $ht$, where, as explained, $h$ now plays the role of the SM-like 
Higgs state. Under these assumptions,
the ATLAS search in Ref.~\cite{atlas-bound} is the most constraining one and
 excludes a heavy $T$ quark with mass lower than $\approx 640$ GeV
at the 95\% Confidence Level (CL). This lower limit can, however, be weakened 
down to $\approx 350$ GeV if $T$ couples to first and
second generation quarks as well \cite{Aad:2012bt}. This is certainly 
a possible model construction, however, in our case, we
do not pursue this in any detail, as such additional interactions would not 
enter the Higgs  boson observables which we intend to
study.  We are nonetheless entitled to scan on $m_T$ starting 
from such low $m_{\rm VLQ}$ values.
%It is instructive to compare the above constraint to results of direct
%experimental searches for up-type singlet vector-like quarks.
%Interestingly, the most severe bound on $T$ in the renormalizable model;
%and assuming dominant but small $T$ mixing with the top,
%$m_{T}>640$~GeV~\cite{atlas-bound} is given by the ATLAS
%experimental search using the $T \to t h$ decay signature.
%However, if $T$ does not dominantly decay to third generation
%quarks (but instead to the first two generations), the current
%direct search constraints are relaxed dramatically (c.f.~\cite{Aad:2012bt})
%and $m_{T} \simeq 300$~GeV becomes a possibility.
In our 2HDM+VLQ construct, if we assume $m_A, m_H, m_{H^\pm}< m_T-m_t$,  then also the
$T\to tH$, $T\to tA$ and $T\to bH^\pm$
decays open up, alongside $T\to th$. 
%
%We do expect  that  Run 2
%data will eventually have sensitivities to such scenarios, which may   
%become testable as the luminosity increases, up to VLQ
%masses of order 1.5 TeV or so, as demonstrated
%in a similar study assuming  chiral
%quarks within a 2HDM \cite{Bar-Shalom:2016ehq,BarShalom:2012ms}. 
%
Finally, one should recall 
that $T$ production at the LHC is substantial, in both
the QCD induced pair production channel (dominant at low $m_T$) and 
the EW mediated single production channel (dominant at high $m_T$).

%In summary, we will scan the mass range 350 GeV 
%$\leq m_{\rm VLQ}\equiv m_T\leq$ 1.5 TeV.
%=======================================
\subsection{Constraints on the $t$-$T$ mixing}
{{In this section, we will show that $t$-$T$ mixing can be constrained 
both from EW  Precision Observables (EWPOs) and from recent LHC data on the $\approx 125$ 
GeV Higgs boson.}}
%=======================================
\begin{figure}
\begin{centering}
\includegraphics[scale=0.75]{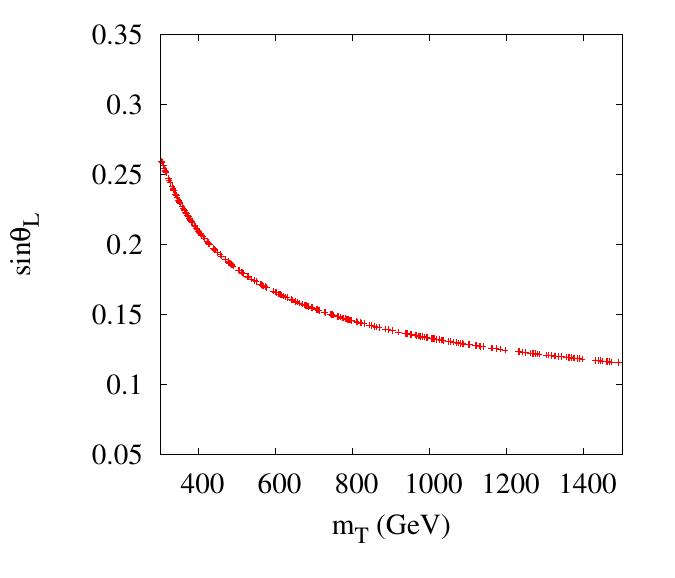}
\end{centering}
\caption{Upper limit at $95\%$ CL on  the  (left-handed) mixing angle
as a function of the $T$ quark mass in the 2HDM-II with an
up-like VLQ.}
\label{fig:1}
\end{figure}
%=======================================
In general, when the new physics scale is much larger than the
EW  scale, virtual effects of the new
particles in loops are expected to contribute to
the EWPOs that have been precisely 
measured at LEP1, LEP2, SLC and Tevatron.
These EWPOs are known as the oblique parameters $S, T$
and $U$ \cite{Peskin:1991sw} and can be used to put constraints on new
physics. In our case, the mixing between $t$-$T$ will generate
 couplings between the SM gauge bosons and the new VLQ, $T$, which will
induce contributions to $S$ and $T$ \cite{Lavoura:1992np}.

We have computed the extra contributions of the VLQ to
$\Delta T=T-T_{\rm SM}$ and $\Delta S=S-S_{\rm SM}$  by implementing the model
into the FeynArts \cite{Hahn:2000kx}, FormCalc \cite{Hahn:1998yk,FC-LT} and
 LoopTools \cite{vanOldenborgh:1990yc,LT} packages,
which are used to calculate the required
 gauge boson self-energies. In fact, in our case, the extra contribution to
$\Delta \{T,S\}$ can be cast into pure 2HDM and VLQ parts, 
such that $\Delta \{T,S\}=\Delta \{T,S\}_{\rm 2HDM} + \Delta \{T,S\}_{\rm VLQ}$.
In the present work, we focus on the decoupling limit where
$m_A = m_H = m_{H^\pm}\gg m_Z$ and $\sin(\beta-\alpha)=1$, or slight departures from it,
which leads to  $\Delta T_{\rm 2HDM} =0$ and $\Delta S_{\rm 2HDM} = 0$. We are then left only with
the extra contribution of the VLQ. A straightforward calculation yields
\begin{eqnarray}
\Delta T_{\rm VLQ} &=& \frac{3 m^2_t s^2_L }{16 m^2_W \pi (-1 + r) s^2_W} \left((-1 +
r)(-2 + (1 + r) s^2_L) - 2 r c^2_L \log r\right)\,\,\,{\rm with}\,\,\,
r = \frac{m_{T}^2 }{m_t^2}, \\
\Delta S_{\rm VLQ} &= &\frac{1}{12 m_Z^2\pi}
 \{-m_t^2 s_L^2 + m_T^2 (s_L^2 + 32 s_W^2 c_W^2) +
     s_L^2 (m_Z^2 (10 - 9 s_L^2) - 6 m_t^2 c_L^2 + 6 m_T^2 c_L^2)
       A_0[m_t^2]\nonumber\\ && +
     (s_L^2 (6 m_t^2 c_L^2 - 6 m_T^2 c_L^2 +
          m_Z^2 (-10 + 9 s_L^2)) - 32 m_Z^2 s_W^2 c_W^2) A_0(m_T^2)
\nonumber \\ &&
   + 3 m_t^2 s_L^2 (10 - 3 s_L^2) B_0(0, m_t^2, m_t^2)
     - 18 m_t^2 s_L^2 c_L^2 B_0(0, m_t^2, m_T^2)
\nonumber\\ &&
 -
     m_T^2 (12 s_L^2 + 9 s_L^4 - 32 s_W^2 c_W^2) B_0(0, m_T^2, m_T^2)
     \nonumber\\ && + 2 s_L^2 (m_Z^2 (2 - 3 s_L^2) + m_t^2 (-14 + 3 s_L^2))
       B_0[m_Z^2, m_t^2, m_t^2] \nonumber\\ &&
+
     6 ((m_t^2 - m_T^2)^2 + (m_t^2 + m_T^2) m_Z^2 - 2 m_Z^4) s_L^2 c_L^2
       B_0[m_Z^2, m_t^2, m_T^2]\nonumber \\ &&  +
     2 s_L^2 (m_Z^2 (4 - 3 s_L^2) + m_T^2 (8 + 3 s_L^2)) B_0[m_Z^2, m_T^2, m_T^2]\},
\end{eqnarray}
where the $A_0$ and $B_0$ functions are the standard Passarino-Veltman ones used in the convention of LoopTools \cite{LT}.
Note that our results agree numerically with Ref.~\cite{Sally}.
Taking the above analytical expressions into account,
our model will remain viable as long as $\Delta T_{\rm VLQ}$
 and  $\Delta S_{\rm VLQ}$ are compatible with the latest 
extracted values \cite{pdg} which are given by
\begin{eqnarray}
 \Delta T = 0.1 \pm  0.07,\quad \quad \quad
 \Delta S = 0.06 \pm  0.09,
 \end{eqnarray}
where a correlation coefficient  $\rho=+0.91$ and $ \Delta U =0$ have been
used. We thus perform a random scan on the $s_L$ and $m_T$ parameters
imposing compatibility with $\Delta T$ and $\Delta S$ at
95$\%$ CL, which yields a constraint on
$s_L$ as a function of the VLQ mass, $m_T$, as shown in {Fig.~\ref{fig:1}}.
One can see that the constraint on the mixing is, e.g.,  $|s_L|\leq 0.25$
for $m_T=350$ GeV and $|s_L|\leq 0.12$
for $m_T=1$ TeV. For a heavy VLQ, the constraint on the mixing is more severe
 and is mainly coming from $\Delta T$ which contains a large logarithm of $m_T^2/m_t^2$.
%%%%%%%%%%%%%%%%%%%%%%%%%%%%%%%%%%%%%%
%=======================================
\subsection{Constraints from $B\to X_s\gamma$ }
In addition to the EWPOs constraints studied above,  the penguin induced $b\to s \gamma$ decay is also sensitive to new physics. The current experimental value is BR$(\bar B \to X_s \gamma)^{\rm exp}=(3.32 \pm  0.15) \times 10^{-4}$, for $E_\gamma > 1.6$ GeV~\cite{Amhis:2016xyh}, and the  SM prediction with Next-to-Next-to-Leading Order (NNLO) QCD corrections  is  BR$(\bar B \to X_s \gamma)^{\rm SM}=(3.36 \pm  0.23) \times 10^{-4}$~\cite{Czakon:2015exa,Misiak:2015xwa}.  Since the SM result is close to  the experimental data, $\bar B \to X_s \gamma$ will give a strict bound on  new physics effects. 
The effective Hamiltonian arising from the $W^\pm$ and $H^\pm$ bosons for $b\to s \gamma$ at the $\mu_b = 4.8$ GeV scale can be written as:
\begin{align}
{\cal H}_{b\to s \gamma} = -\frac{4G_F}{\sqrt{2}} \sum_{i=t,T}V^*_{is} V_{ib} \left(C^i_{7\gamma} (\mu_b)O_{7\gamma}  + C^i_{8\gamma} (\mu_b)Q_{8G} \right)\,, 
\end{align}
where the EM and gluonic dipole operators are given as:
 \begin{equation}
 O_{7\gamma} = \frac{e}{16\pi^2} m_b \bar s \sigma^{\mu\nu} P_R b F_{\mu\nu}\,, \  O_{8G} = \frac{g_s}{16\pi^2} m_b \bar s_\alpha \sigma^{\mu\nu}  T^a_{\alpha \beta} P_R b_\beta G^a_{\mu\nu}\,.
 \end{equation}
Here, 
$C_{7\gamma}(\mu_b)$ and $C_{8G}(\mu_b)$ are the Wilson coefficients at the $\mu_b$ scale and their relations to the initial conditions at the high energy scale $\mu_H$ (needed to describe the evolution from such a high scale down to the lower energy $\mu_b$ via the matching scale $\mu_0$~\cite{Borzumati:1998tg}) are through Renormalization Group Equations (RGEs). The NLO~\cite{Ciuchini:1997xe,Borzumati:1998tg,Borzumati:1998nx} and NNLO~\cite{Hermann:2012fc} QCD corrections to  $C_{7\gamma}(\mu_b)$ and $C_{8G}(\mu_b)$ in the 2HDM-II have been calculated.  Based on the $C^{\rm SM}_{7\gamma}(\mu_b)$ value extracted in~\cite{Blanke:2011ry}, we get $C^{\rm SM}_{7\gamma}(\mu_b) \approx -0.310$ when BR$(\bar B \to X_s \gamma)^{\rm SM}=3.36 \times 10^{-4}$ is applied. 
In order to study the influence of the $b\to s \gamma$ process on the 2HDM+VLQ model, we follow the approach  in~\cite{Misiak:2015xwa} and split the BR$(\bar B \to X_s \gamma)$ as follows:
 \begin{align}
  {\rm BR}(\bar B \to X_s \gamma) \times 10^{4} \approx (3.36 \pm 0.23) -8.22\, Re(C^{i}_{7\gamma}) -1.99\, Re(C^{i}_{8G})\,,
  \end{align}
 where $C^{i}_{7\gamma, 8G}$ are the Wilson coefficients at the $\mu_H = m_{H^\pm}$ scale (the matching scale is $\mu_0 \sim m_t$ at which the heavy particles are decoupled~\cite{Misiak:2015xwa}) 
%{\textcolor{blue}{Too many scales, $\mu_0, \mu_b, \mu_H$ and $\mu_W$, the latter being undefined: it needs clearing up!}},
, wherein the quadratic $C^{i}_{7\gamma,8G}$ terms  are ignored due to the requirement of $C^{i}_{7\gamma,8G}<1$.  Using the current experimental value, the bound on $C^{i}_{7\gamma,8G}$ is:
    \begin{equation}
 8.22 Re(C^{i}_{7\gamma}) + 1.99 Re(C^{i}_{8G}) \approx  0.04 \pm 0.28.  \label{eq:C7bound}
 \end{equation}
According to the charged Higgs  interactions, %in Eq.~(\ref{eq:btoqpp})
 the $H^\pm$  contributions  coming together with $t$ and $T$ to $C^{i}_{7\gamma,8G}$ are expressed as~\cite{Borzumati:1998tg}: 
%%%%%%%%
\begin{align}
 C^{i}_{7\gamma} & = f_{1\gamma}(x_i)/\tan^2\beta + f_{2\gamma}(x_i) \,, \nonumber \\
C^{i}_{8G} &=   f_{1G}(x_i)/\tan^2\beta + f_{2G}(x_i), 
 \label{eq:CCH7}
 \end{align}
with $x_i = m^2_i/m^2_{H^\pm}$ and $i = t, T$. The form factors are given in Appendix B.

\begin{figure}
\begin{centering}
\includegraphics[scale=0.565]{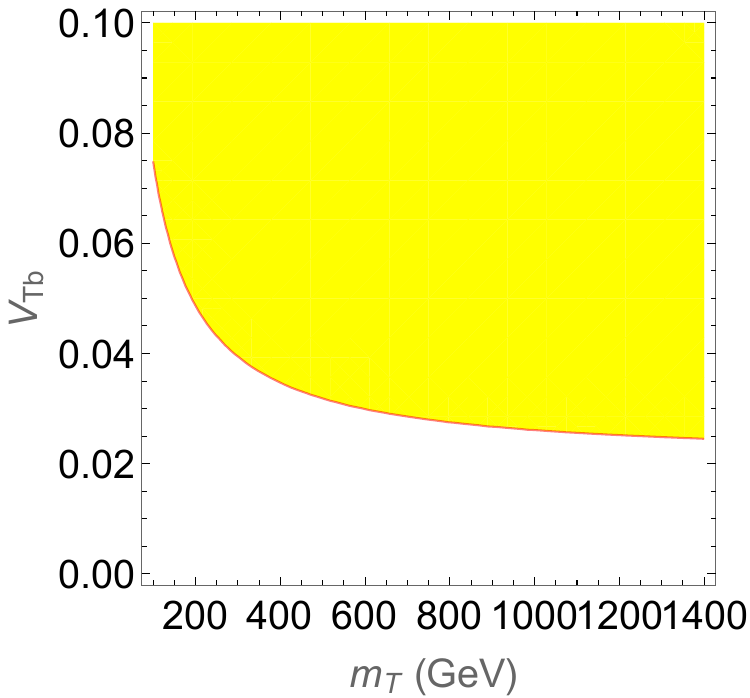}
\includegraphics[scale=0.53]{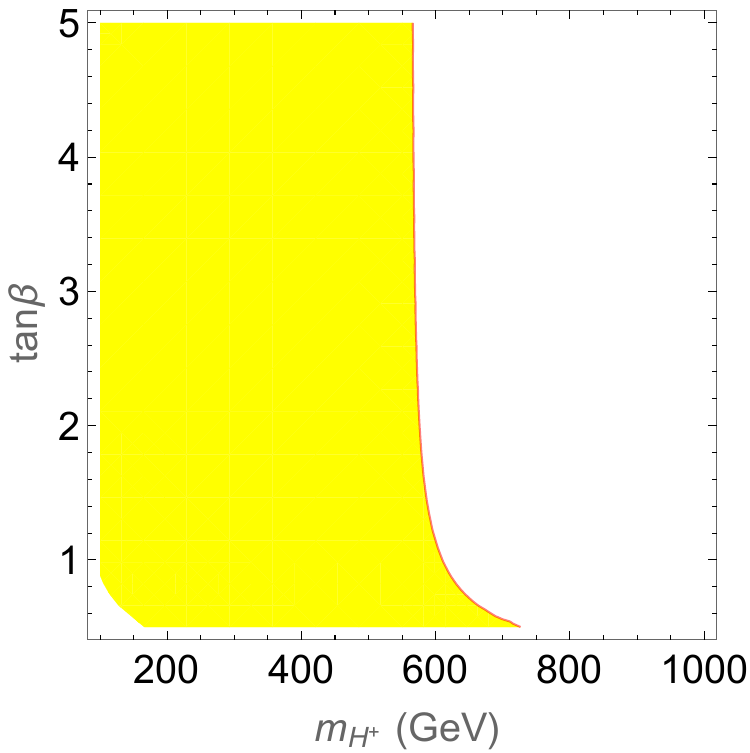}
\end{centering}
\caption{ Excluded regions (yellow color) at $95\%$ CL from BR($\bar{B}\to X_s\gamma$) in the 2HDM+VLQ. 
Left panel: Limit on  $V_{Tb}$ as a function of $m_T$ with $m_{H^\pm} = 600$ GeV.
Right panel: Limit on the charged Higgs mass with $m_T = 400$ GeV and $V_{Tb} = 0.025$.}
\label{fig:Exclude_BR}
\end{figure}

%{\textcolor{blue}{Appendix C is cited before Appendix B, which is not mentioned at all in the text}} 
Fig. \ref{fig:Exclude_BR} shows the limits on $V_{Tb}$ (left) and $m_{H^\pm}$ (right) in the 2HDM+VLQ. We learn that BR$(b\to s\gamma)$ excludes $V_{Tb}\ge 0.03$ regardless of the value of $m_T$ used. It further appears from the right panel of Fig. \ref{fig:Exclude_BR} that large $\tan\beta$ is not excluded by current data and a lower bound $m_{H^\pm}\ge $ 600 GeV is obtained similar to the case of the standard 2HDM-II, (Recall that a lower bound $m_{H^\pm}\ge $ 580 GeV was obtained in \cite{Misiak:2017bgg} for  this case.)
%=======================================
%%%%%%%%%%%%%%%%%%%%%%%%%%%%%%%%%%%%%%%%%%%%%%%%%%%%%%
\subsection{Constraints from LHC data}
\begin{figure}[t!]
\hspace*{-1.75cm}
\includegraphics[width=0.46\textwidth]{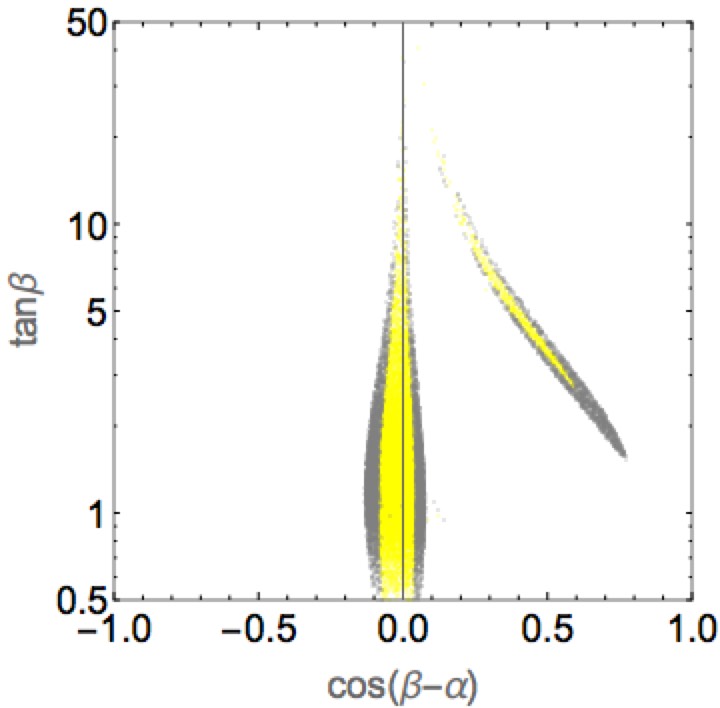}
\includegraphics[width=0.46\textwidth]{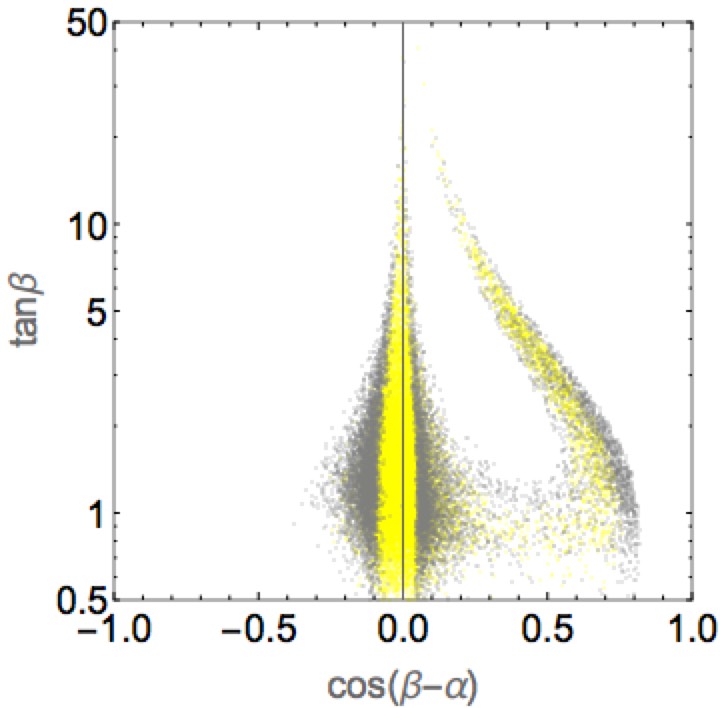}
%\vspace{-18cm}
\caption{Constraints from the LHC Higgs data on the parameter space of the ordinary 2HDM (left) and 2HDM+VLQ  (right). 
We show the 95$\%$ CL region in the $(\cos(\beta-\alpha), \tan\beta)$ plane. The gray (yellow) regions were obtained using Higgs Run 1(2) data. We have varied  $|s_L| < 0.20$ and 400 GeV$  < m_T < $ 1000 GeV and set $y_T = 4\pi$. }
\label{fig:LHC_constraint}
\end{figure}

The  couplings of the SM-like Higgs are sensitive to the parameters $\cos(\beta-\alpha)$ and $\tan\beta$. Therefore, the LHC data on the 125 GeV Higgs boson can give  strong constraints on these. In Fig.~\ref{fig:LHC_constraint},
we show the constraints on the ordinary 2HDM (left) and 2HDM+VLQ (right) using Higgs data from Run 1 (gray) and Run 2 (yellow) at 95$\%$ CL. The bounds on $\cos(\beta-\alpha)$ are much more stringent for the ordinary 2HDM-II, where the SM-like coupling region of the 125 GeV Higgs forces $|\cos(\beta-\alpha)| < 0.14$, and increasingly more stringent for larger $\tan\beta$. However, in the so-called `wrong sign' Yukawa coupling region of the 125 GeV Higgs state, we find 
%{\textcolor{blue}{It should be explained what this is or a reference given}}, 
$|\cos(\beta-\alpha)| < 0.45$. By varying $|s_L| < 0.20$ and 400 GeV$< m_T < $ 1000 GeV and setting $y_T = 4\pi$, the situation in the 2HDM+VLQ (right panel) is quite different for low $\tan\beta$, where $|\cos(\beta-\alpha)| < 0.4$.\\
%%%%%%%%%%%%%%%%%%%%%%%%%%%%%%%%%%%%%%%%%%%%%%%%%%%%%%
%%%%%%%%%%%%%%%%%%%%%%%%%%%%%%%%%%%%%%

Another constraint, this time on the mixing, $s_L$, comes from the contribution of
the VLQ to the di-photon event rate of the $\approx 125$ GeV SM-like Higgs boson.
The modified top quark coupling to this Higgs boson
and the presence of an additional heavy quark can
 impact loop induced Higgs decays, namely, $h\to gg$,
 $h\to\gamma\gamma$ and $h\to Z\gamma$.
The relevant partial decay widths are 
given by\footnote{The analytical expressions for $\phi\to gg$ decays
are easily obtainable from those for $\phi\to \gamma\gamma$. Similarly, for $gg\to \phi $ production.}
%===========================================
\begin{eqnarray}
\Gamma(\phi\to \gamma\gamma) &=& \frac{G_F \alpha^2 m^3_{\phi}}{128\sqrt{2}\pi^3}
\left|\frac{4}{3}\sum_{f=t,T}\kappa^{\phi}_{ ff}
A^{\phi}_{1/2}(\tau_{f}) +
\kappa^{\phi}_{WW} A^{\phi}_{1}(\tau_{W}) +
g^{\phi}_{H^\pm H^\mp}\frac{m^2_W}{2 c^2_W m^2_{H^\pm}}A^{\phi}_{0}(\tau_{H^\pm}) \right|^2,~~~\\
%===========================================
%===========================================
\Gamma(\phi\to Z\gamma) &=& \frac{G^2_F m^2_W \alpha m^3_{\phi}}{64\pi^4}\left(1-\frac{m^2_Z}{m^2_{\phi}}\right)^3 
\Bigg|\sum_{f=t,T}\kappa^\phi_{ff}\frac{2 n_f Q_f (I^3_f - 2s^2_W Q_f)}{c_W}
A^{\phi}_{1/2}(\tau_{f},\lambda_{f}) \\\nonumber &+& \sum_{f,f'=t,T} \frac{m_{f(f')}}{m_W}\sum_{m=L,R}\kappa^\phi_{ff'}g^m_{ff'}
A^{\phi}_{1/2}(\tau_{f},\lambda_{f'}) +  \kappa^{\phi}_{ WW} A^{\phi}_{1}(\tau_{W}) +
g^\phi_{H^\pm H^\mp}\frac{m^2_W}{m^2_{H^\pm}} \frac{c_{2W}}{c_W} A^{\phi}_{0}(\tau_{H^\pm}) \Bigg|^2\\
%============================================
%===========================================
\Gamma(A\to \gamma\gamma) &=& \frac{G_F \alpha^2
  m^3_{A}}{128\sqrt{2}\pi^3}\left|\frac{4}{3} \sum_{f=t,T}\kappa^A_{ff} A^{A}_{1/2}(\tau_{f})\right|^2,\\\nonumber
%============================================
\Gamma(A\to Z\gamma) &=& \frac{G^2_F m^2_W \alpha m^3_{A}}{16\pi^4}\left(1-\frac{m^2_Z}{m^2_{A}}\right)^3 \\\nonumber &&
\Bigg|\sum_{f=t,T}\kappa^A_{ff}\frac{2 n_f Q_f (I^3_f - 2s^2_W Q_f)}{c_W} A^{A}_{1/2}(\tau_{f},\lambda_{f}) +  \sum_{f,f'=t,T} \frac{m_{f(f')}}{m_W}\sum_{m=L,R}\kappa^A_{ff'}g^m_{ff'}
A^{A}_{1/2}(\tau_{f},\lambda_{f'}))\Bigg|^2 ,\\\nonumber
%===============================================
\end{eqnarray}
where $\phi = h(\equiv h_{\rm SM}$) or $H$ are the CP-even Higgs bosons of
the  2HDM, with $\kappa^{h(H)}_{WW} = \sin(\beta-\alpha) (\cos(\beta-\alpha))$.
The relevant loop functions can be found 
in, e.g., Refs.~\cite{Spira:1997dg,Djouadi:2005gi}.
Clearly, the charged Higgs boson contributions are smaller compared to the fermionic ones.
Even for large $g_{\phi H^\pm H^\mp}$, the charged Higgs effects are still
negligible, henceforth, we neglect these.

The relevant modifications to the signal strength $\mu^h_{\gamma\gamma}$
(a function of the production cross-sections and  decay BRs) are defined in our scenario as
\begin{equation}\label{eq:muAA}
\mu^h_{V\gamma}\equiv\frac{\sigma^{\rm VLQ}(pp\to \phi)}{
\sigma(pp\to h_{\rm SM})}\times
\frac{{\rm BR}^{\rm VLQ}(\phi\to V\gamma )}{{\rm BR}(h_{\rm SM} \to V\gamma)},
\quad \quad V = Z,\gamma,\quad \phi = H,A~(\rm{summed~over}).
\end{equation}
These come from the presence of an additional VLQ in the loops as well as from the
modification of the $ht\bar{t}$ coupling for both 
Higgs production ($gg\to \phi$)
and Higgs decay ($\phi \to V\gamma$). The theoretical value for
$\mu^h_{\gamma\gamma}$ will depend on $m_T$, $s_L$ as well as  the new
Yukawa $y_T$. However, in the decoupling limit $\cos(\beta-\alpha)=0$,
the dependence of $ht\bar{t}$ and $hT\bar{T}$ on $y_T$ cancels due to a
factor $\cos(\beta-\alpha)=0$. What then remains is solely an $m_T$
and $s_L$ dependence, at least in the $V=\gamma$ case. The formula in Eq. (\ref{eq:muAA}) holds for the $Z\gamma$ case as well, wherein, however, the
role of the $T$ loops can be altered significantly relative to that of the others by the additional degree of freedom
carried by the $ZT\bar T$ vertex (unlike the case of the $\gamma T\bar T$ one, which is fixed by the Ward identity).
Further, unlike the case of $\gamma\gamma$, $Z\gamma$ also benefits from non-diagonal loop transitions wherein
the vertices $H,A t\bar T$ and $Z t \bar T$ (and c.c.) are involved.
These differences between the two decay channels will play a key role in the remainder of our analysis.

\begin{figure}[t!]
\hspace*{-1.75cm}
\includegraphics[width=1.1\textwidth]{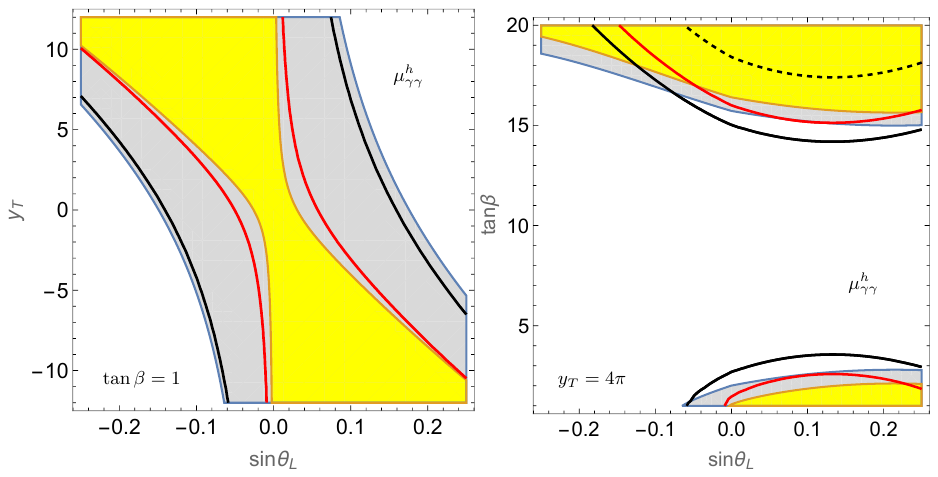}
\vspace{-18cm}
\caption{Allowed regions for $\mu^h_{\gamma\gamma}$ in the 2HDM+VLQ at 95$\%$CL of Higgs data Run 1 (gray) and Run 2    
  (yellow)
over the ($s_L, y_T$) plane (left) with $\tan\beta = 1$ and over the  ($s_L, \tan\beta$) plane (right) with $y_T = 4\pi$. 
The other parameters are fixed to $m_T = 1$ TeV, $\cos(\beta-\alpha) = 0.05$.  The contour plots correspond to $\mu^h_{Z\gamma} = 1$ (dashed black), 1.5 (solid red) and 2 (solid black).}
\label{fig:3}
\end{figure}

The effects of a new heavy quark, $T$, have direct consequences for the signal strengths of the SM-like Higgs boson.
In Fig.~\ref{fig:3}, we illustrate a contour plot for
$\mu^h_{\gamma\gamma}$ over the ($s_L$, $y_T$) plane. The dashed black, 
solid black and solid red contours  capture $\mu^h_{Z\gamma} = 1$, 1.5 and 2, respectively.
The three contours fall within 1$\sigma$ of the
ATLAS and CMS measured value of $\mu^h_{\gamma\gamma}= 1.10\pm 0.23~({\rm
  stat})\pm 0.22~({\rm syst})$. One can therefore conclude that this LHC 
 constraint is less stringent than the oblique parameters previously 
 discussed. {{However, once the $\mu^h_{\gamma\gamma}$ measurement improves with Run 2 data from the 
 LHC and reaches the level of 10\% or less deviation from the SM value, then the $\approx 125$ GeV di-photon
 event constraints  will be more stringent.}} The pattern of $\mu^h_{Z\gamma}$ is also given. It is remarkable that, for $\mu^h_{\gamma\gamma}$ compatible with LHC data at the $\pm2\sigma$ level, $\mu^h_{Z\gamma}$ can see an enhancement up
to a factor of nearly 2.
%%%%%%%%%%%%%%%%%%%%%%%%%%%%%%%%%%%%%%%%%%%%%%%%%%%%%%
%%%%%%%%%%%%%%%%%%%%%%%%%%%%%%%%%%%%%%
%%%%%%%%%%%%%%%%%%%%%%%%%%%%%%%%%%%%%%
%\vspace{-1cm}

%=================================================
\section{Confronting the 2HDM+VLQ with LHC data}
\label{sec:III}
In order to explain the LHC data in the framework of our 2HDM+VLQ construct, we consider the 
$gg\to \phi$ and $\phi\to \gamma \gamma$ (with $\phi = H$ and $A$) processes where
the contribution of all the quarks including $T$ is considered. In the
SM, only the top-quark loop gives a significant fermionic contribution. Besides the
top-quark, the new VLQ state $T$ can also contribute in the 2HDM+VLQ case, for 
both the SM-like and the other heavy Higgs production modes in addition to their decays into di-photons. 

We start by assuming that we are in the
alignment  limit of the light Higgs boson $h$, $\cos(\beta -\alpha) = 0$, wherein the heavier CP-even
Higgs boson, $H$, decays to $W^+W^-$ and $ZZ$ vanish at the tree level, which is
consistent with current Higgs boson searches. (Needless to say, the $A$ state
cannot directly couple to pairs of SM massive gauge bosons in presence of CP
conservation). We also assume that the heavy Higgs states are degenerate, $m_H =
m_A$, which is favored by satisfying theoretical bounds
such as vacuum stability, perturbativity and allowed
by EWPOs \cite{Olive:2016xmw}.

%%%%%%%%%%%%%%%%%%%%%%%%%%%%%%%%%%%%%%%%%%%%%%%%%%%%%%
\begin{figure}[h!]
\hspace{-1.69cm}\includegraphics[width=1.1\textwidth]{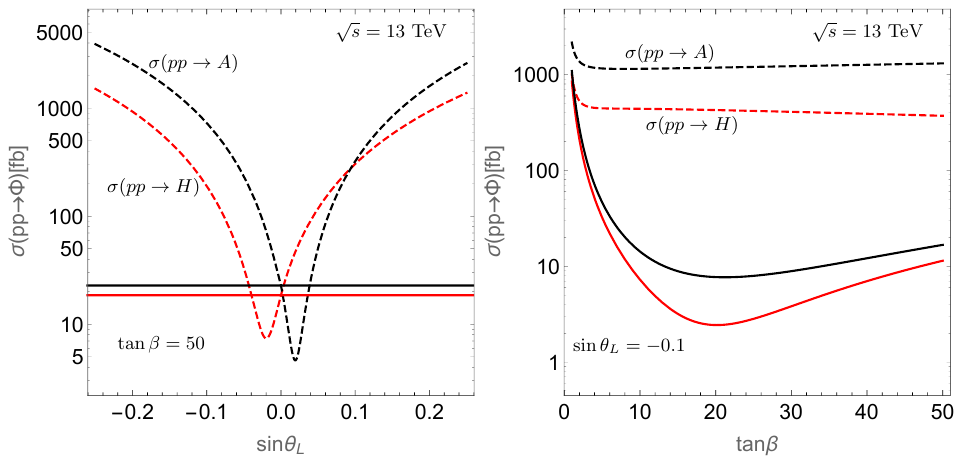}
\vspace{-18cm}
\caption{Heavy Higgs production $\sigma(pp\to A, H)$ in fb at the LHC with $\sqrt s=13$ TeV as a function of $\sin\theta_L$ (left) and $\tan\beta$ (right) for $m_H = m_A = 700$ GeV in the alignment limit of the 2HDM+VLQ. We fix $y_T = 2\pi$, $ m_T = 700$ GeV and $\cos(\beta-\alpha) = 0$. The solid (black and red) lines correspond to 2HDM without VLQ. 
}
\label{fig:4}
\end{figure}
%%%%%%%%%%%%%%%%%%%%%%%%%%%%%%%%%%%%%%%%%%%%%%%%%%%%%%
The cross-section for this process is given by
\begin{equation}\label{eq:sigma}
\sigma^{\rm VLQ}(pp\to \phi) = \sigma(pp\to h_{\rm SM})\times
\frac{\Gamma^{\rm VLQ}(\phi\to gg)}{\Gamma(h_{\rm SM} \to gg)}, 
\quad \quad \phi = H,A,
\end{equation}
where $\Gamma^{\rm VLQ}(\phi\to gg)$ is the $gg$ width of the SM augmented by the extra VLQ loop contribution. The SM 
Higgs cross-section is taken from the Higgs working group study of
Ref. \cite{HWG}. 
In our 2HDM+VLQ scenario, the cross-section can be enhanced from the additional VLQ loop, which introduces the
 $hT\bar{T}$ coupling which can be large. Its sign is
such that it can enable constructive interference
with the top quark loop.
Furthermore, the BR$^{\rm VLQ}(H,A\to\gamma\gamma)$ can
overall be enhanced as well  
through a similar dynamics, though it should be
recalled here that the dominant loop is due to $W^\pm$'s which typically have
an opposite sign to the $t$ and $T$ loops, owing to the different spin
statistics.
%%%%%%%%%%%%%%%%%%%%%%%%%%%%%%%%%%%%%%%%%%%%%%%%%%%%%%
%%%%%%%%%%%%%%%%%%%%%%%%%%%%%%%%%%%%%%
In Fig.~\ref{fig:4}, fixing $m_H = m_A = 700$ GeV,
we present the dependence of  $\sigma(pp \to H, A)$  upon $s_L$ for $\tan\beta = 50$
while the right panel shows the same quantity as a function of $\tan\beta$ with fixed mixing angle $s_L=-0.1$. 
Both panels are for a large Yukawa of the new top, $y_T=2\pi$ with $ m_T = 700$ GeV. Here, 
$\tan\beta$ is required to be smaller than 20 for $m_H = m_A = 700$ GeV.
It is clear from Fig.~\ref{fig:4} that, away from the $s_L\approx 0$ limit (left frame), the $pp\to H,A$ process can differ by two orders of magnitude compared to the ordinary 2HDM-II. 

%%%%%%%%%%%%%%%%%%%%%%%%%%%%%%%%%%%%%%%%%%%%%%%%%%%%%%
\begin{figure}[t!]
\hspace*{-1.1cm}\includegraphics[width=1.1\textwidth]{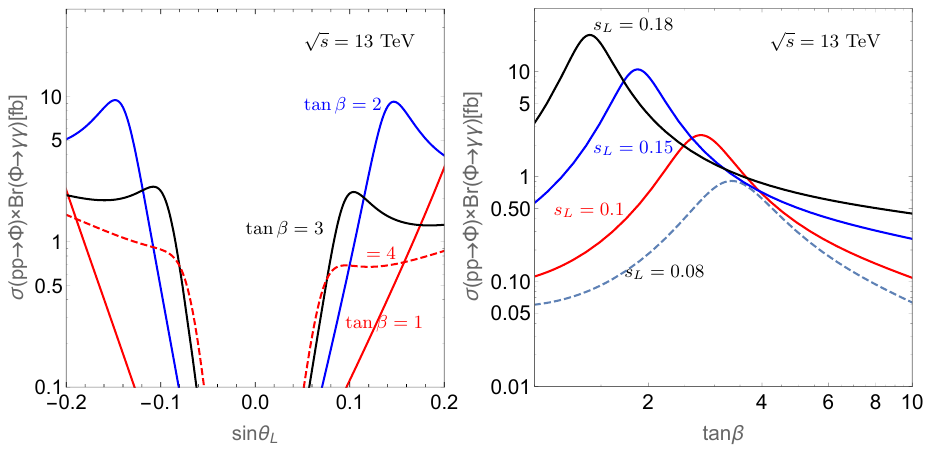}\\
\vspace{-18cm}
%\hspace*{0.5cm}\includegraphics[width=0.35\textwidth]{zoom.pdf}
\caption{Cross sections of $(pp\to H, A \to \gamma\gamma)$ in fb  
  in the 2HDM+VLQ as a function of $s_L$ for different values of $\tan\beta$ (left) and $\tan\beta$ for different values of $s_L$ (right). Here, $\cos(\beta - \alpha) = 0$, $m_H = m_A = 700$ GeV, $m_T = 700$ GeV and $y_T = 4\pi$.}
\label{fig:5}
\end{figure}
%%%%%%%%%%%%%%%%%%%%%%%%%%%%%%%%%%%%%%
%%%%%%%%%%%%%%%%%%%%%%%%%%%%%%%%%%%%%%
In the left panel of Fig.~\ref{fig:5} we present the
dependence of  $\sigma(pp \to\phi)\times 
{\rm BR}(\phi\to\gamma\gamma)$  upon the mixing angle $s_L$ for $\tan\beta=1,2,3,4$ while the
 right panel  shows the same quantity as a function of $\tan\beta$ with fixed mixing angles $s_L$. Both panels are for a large Yukawa of the new top, $y_T=4\pi$.
In this illustration, we limit ourselves to low $\tan\beta$ values which are favored by
$\tau\tau$  LHC data and also because of perturbative unitarity coming from the 2HDM scalar potential. 
We emphasize  that, 
in the decoupling limit which we consider, the $W^\pm$ loop in $H\to \gamma\gamma$ 
 vanishes since it is proportional to $\cos(\beta-\alpha)\approx 0$ while 
$A\to \gamma\gamma$ has no $W^\pm$ loop at all because of the CP-odd nature of  the $A$ state.
Furthermore, for $H,A\to gg$, we are only left with top quarks and VLQ contributions.
{
It is clear from Fig.~\ref{fig:5} that, away from the $s_L\approx 0$ limit (left frame), the $pp\to H,A\to \gamma\gamma$ process can significantly contribute to the high mass di-photon event sample. Hence, the recent studies carried out by ATLAS and CMS have the potential to significantly constrain our model. For example, for $H,A$ masses around 500 GeV, $\tan\beta$ values of 2--3 are not possible, as no particular feature has emerged from the LHC data in the relevant  $m_{\gamma\gamma}$ invariant mass range (hence they are compatible with the SM rates, driven by $q\bar q,gg\to\gamma\gamma$ events).  In fact, the strongest constraints would emerge (right frame) for any $m_A=m_H$ value whenever a loop threshold opens up, whether this is the $t\bar t$ one at 350 GeV 
or the $T\bar T$ one at higher energies
(possibly excluding $\tan\beta\approx4$ values), which may happen through
both Higgs \cite{HHG} and $Z$ \cite{Moretti:2014rka,Jain:2016rhk} boson mediation. 
}

%%%%%%%%%%%%%%%%%%%%%%%%%%%%%%%%%%%%%%%%%%%%%%%%%%%%%%
\begin{figure}[h!]
\hspace*{-1.1cm}\includegraphics[width=1.1\textwidth]{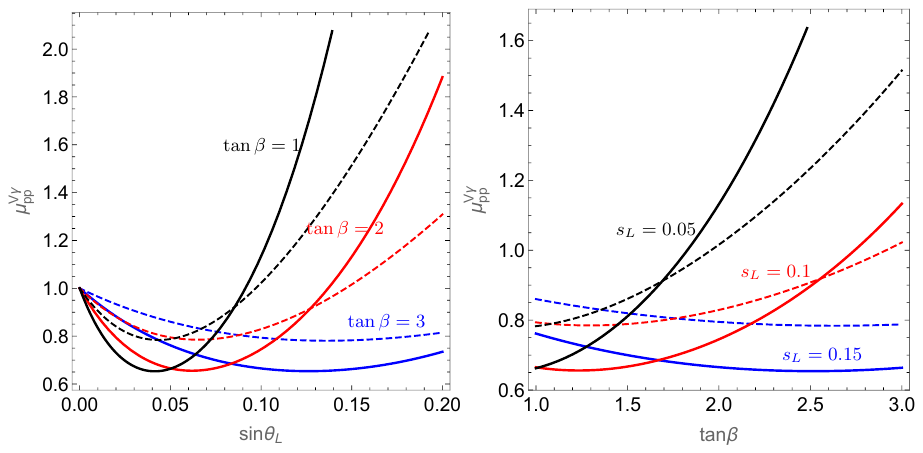}\\
\vspace{-18cm}
%\hspace*{0.5cm}\includegraphics[width=0.35\textwidth]{zoom.pdf}
\caption{Ratios of $\sigma(gg\to H, A \to \gamma\gamma)$ (solid)  and $\sigma(gg\to H, A \to Z\gamma)$ (dashed) 
  in the 2HDM+VLQ over the 2HDM for $m_{\gamma\gamma}=m_{Z\gamma}=750$ GeV
as a function of $s_L$ for different values of $\tan\beta$ (left) and Yukawa coupling $y_T$ (right). Here, $\cos(\beta - \alpha) = 0$, $m_H = m_A = 700$ GeV, $m_T = 700$ GeV.  
Also, we fix, $y_T = -2\pi$.}
\label{figg:9}
\end{figure}
%%%%%%%%%%%%%%%%%%%%%%%%%%%%%%%%%%%%%%
A point we have previously made regarding our 2HDM+VLQ construct is the possibility of $\gamma\gamma$ rates at high invariant masses being compatible with Run 2 data 
with the $Z\gamma$ ones being potentially different from the standard 2HDM-II case. With this in mind, we present  
Fig. \ref{figg:9}, where the inclusive rates of these two channels in the 2HDM+VLQ are shown, divided by the corresponding 
2HDM rates (these correspond to the case of $s_L=0$ and $m_T\to\infty$), i.e.,
\begin{equation}
\mu^{V\gamma}_{pp}\equiv\frac{\sigma(gg\to H, A \to V\gamma)_{\rm 2HDM+VLQ}}
                                                      {\sigma(gg\to H, A \to V\gamma)_{\rm 2HDM        }},~~~~V=Z,\gamma, 
\end{equation}
for low $\tan\beta$ and negative $y_T$. It is clear that substantial differences (up to a factor of two) can exist between the two scenarios, so long
that $s_L$ is sizably different from zero. (Local maxima in the plot correspond to relative sign changes between the VLQ loop contributions and the 2HDM ones). In fact, over most of the possible $s_L$ interval, both $\gamma\gamma$ and $Z\gamma$ in the 2HDM+VLQ depart simultaneously from the ordinary 2HDM-II case. 
%%%
%However, an interesting situation happens in the limit $s_L\to0$,
%i.e., one can see that $\mu^{\gamma\gamma}_{pp} $ and $ \mu^{Z\gamma}_{pp}$ can be significantly different, at the level of a (measurable) $-20\%$ or %so, irrespective of the values of $\tan\beta$ and $y_T$.
%It is also remarkable that this pattern is actually independent of the value chosen for $m_T$ (here being 700 GeV).

{
As pointed out in the introduction, ATLAS and CMS have reported a possible increase in the signal strength of the $t\bar{t}h$ associated production mode in the LHC data. The most recent preliminary results from Run 2 relayed by CMS still show an enhancement of $\mu^{pp}_{t\bar{t}h}$ = 1.5 $\pm$ 0.5 times the SM prediction with an observed significance of 
3.3$\sigma$ compared to the expected one of 2.5$\sigma$ (obtained from combining results of Run 1). Many different final states contribute to this enhancement, but the most significant excesses are observed in multi-lepton final states which probe closely  $t\bar{t}h$ production.
One possibility to explain the excess is that it could be due to the modified Higgs coupling to the SM top quark, resulting in an enhanced $t\bar{t}h$ production. Mixing within the top sector, i.e., between the $t$ and $T$ states, also allows for a sufficiently large enhancement of the $pp\to t\bar{t}h$ rates. Hence, a possibility offered by our scenario could be the potential to explain a $t\bar t h$ enhancement. In this case, rates for the loop-induced processes of the $h$ should remain SM-like, despite the VLQ contributions in our scenario.  
%%%%%%%%%%%%%%%%%%%%%%%%%%%%%%%%%%%%%%
%%%%%%%%%%%%%%%%%%%%%%%%%%%%%%%%%%%%%%%%%%%%%%%%%%%%%%
\begin{figure}[t!]
%\hspace*{-1.1cm}
%\includegraphics[width=1.1\textwidth]{figure9.pdf}\\
\includegraphics[width=0.43\textwidth]{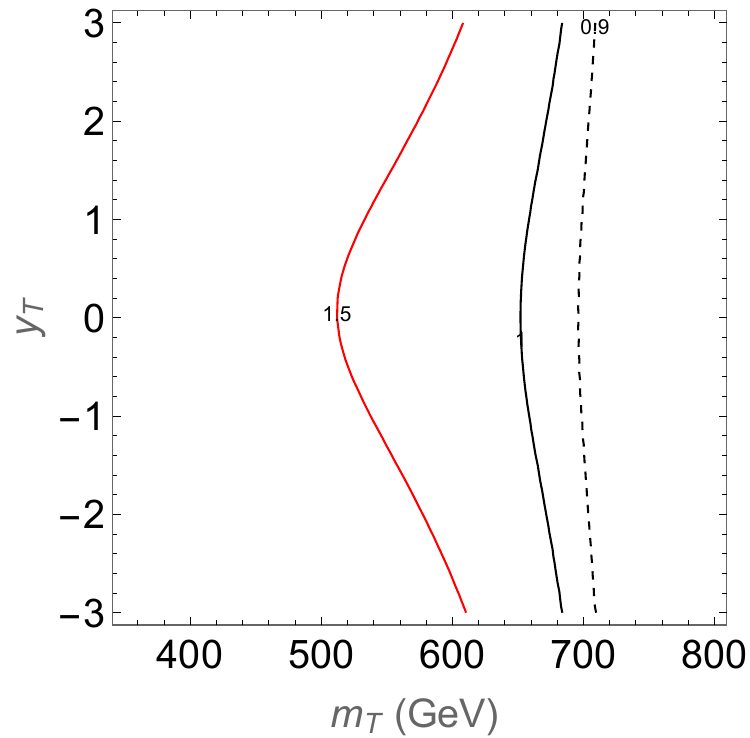}
\includegraphics[width=0.42\textwidth]{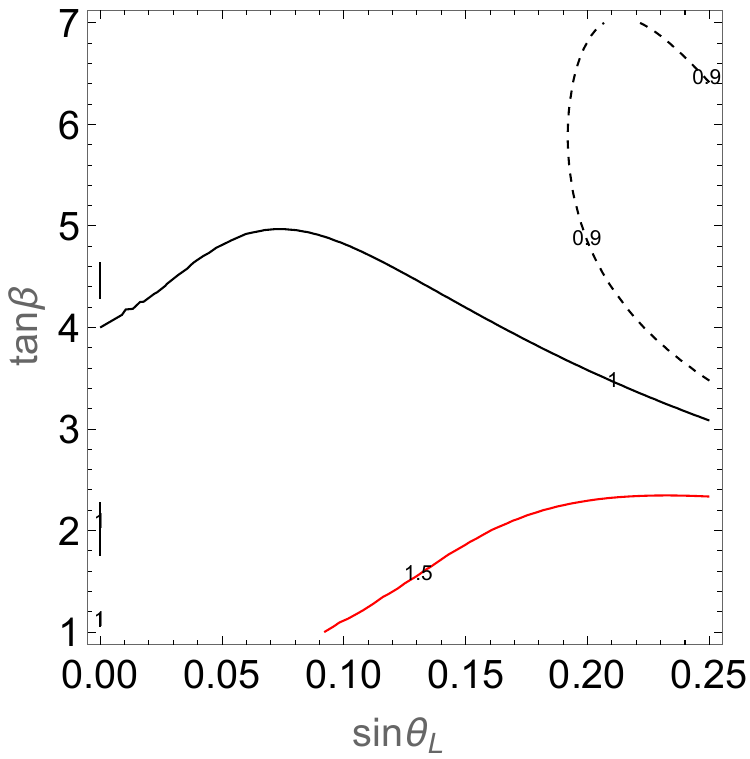}
%\vspace{-18cm}
\caption{Contour plots for the Higgs signal strength $\mu^{pp}_{t\bar{t}h}$ over the ($\sin\theta_L$, $\tan\beta$) plane (right) and over the ($m_T$, $y_T$) plane (left) in the 2HDM+VLQ with $\cos(\beta-\alpha) = 0$, $m_h = 125.5$ GeV, $m_{H} = m_{A}=m_{H^\pm} = 600$ GeV for ($\sin\theta_L$, $\tan\beta$)=(0.22, 1.5) in the left panel and ($m_T$, $y_T$)=(400 GeV, 2)  in the right panel,  at the LHC Run 2. In both plots the red line corresponds to $\mu^{pp}_{t\bar{t}h}=1.5$, the solid black one  to $\mu^{pp}_{t\bar{t}h}=1$ and the dashed black one to $\mu^{pp}_{t\bar{t}h}=0.9$.} 
\label{figg:tth}
\end{figure}
%%%%%%%%%%%%%%%%%%%%%%%%%%%%%%%%%%%%%%
In Fig.~\ref{figg:tth} we show contour plots of $\mu^{pp}_{t\bar{t}h}$ in the ($\sin\theta_L$, $\tan\beta$) plane (right) and ($m_T$, $y_T$) plane (left) in the 2HDM+VLQ given SM-like Higgs couplings. As can be seen from the figure, there is a strong dependence upon both the parameters $\sin\theta_L$ and $\tan\beta$. Clearly, the 2HDM+VLQ can reproduce a higher value of the $t\bar{t}h$ signal strength than in the SM, typically $\mu^{pp}_{t\bar{t}h} \approx 1.5$, for small $\tan\beta$ and  $\sin\theta_L = 0.22$, i.e., a parameter space configuration ideally testable within the experimental range of LHC Run 2 through direct $T$ production.
%%%%%%%%%%%%%%%%%%%%%%%%%%%%%%%%%%%%%
}

%%%%%%%%%%%%%%%%%%%%%%%%%%%%%%%%%%%%%%%%%%%%%%%%%%%%%%%%%%%%%%%%%%%%%%%%%%%
In fact, in the light of a possible explanation of potentially anomalous  $t\bar t h$ data afforded by a heavy top with $m_T\approx 600$ GeV, we end this section with a few comments on the possible production and decay patterns for such a VLQ state, as the ensuing
signatures would be a distinctive feature between a standard 2HDM-II and its
VLQ version. Unlike the case of the SM+VLQ framework where the BR
 of $T\to bW^+$, $T\to tZ$ and $T\to th$ are,  respectively, 50\%, 25\% and 25\%
 for heavy $m_T$, in models with more than one Higgs doublet, several decay
 patterns can appear from the interaction of the new heavy quark 
with the extended Higgs sector, e.g.,  
\begin{eqnarray}
T \quad \to \quad  bW^+, \quad tZ, \quad th, \quad tH, \quad tA, \quad b H^+,
\end{eqnarray}
where the last three cases are unique to a 2HDM sector.
(The partial widths for all these modes are given in  Appendix C). 
%%%%%%%%%%%%%%%%%%%%%%%%%%%%%%%%%%%%%%
\begin{figure}[hptb]%SJDK: uncomment file and vspace
\begin{center}
\includegraphics[width=1.02\textwidth]{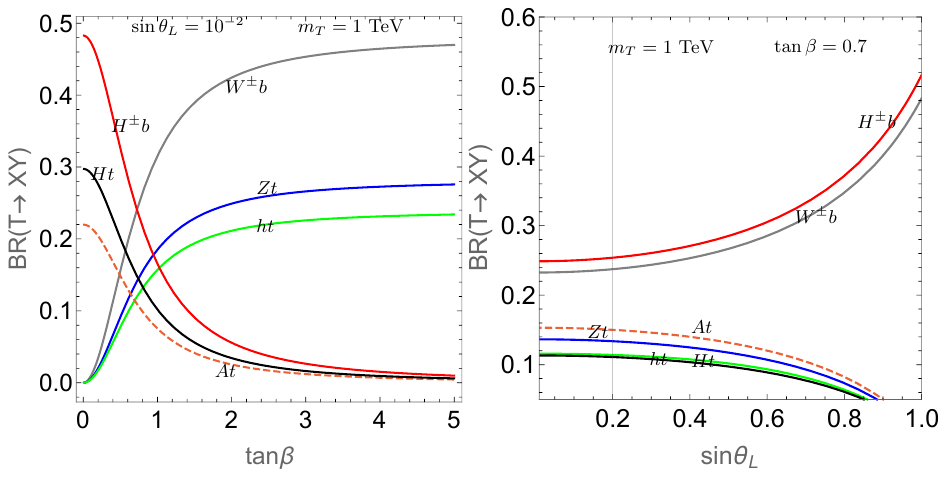}
\end{center}
\vspace{-16cm}
\caption{ Branching ratios of $T$ in the 2HDM+VLQ as a function of $\tan\beta $ (left) and a a function of $\sin\theta_L$ (right) with $y_T = 2$, $m_h = 125.5$ GeV, $m_A = m_H  = 500$ GeV, $m_{H^\pm} = 600$ GeV, $\cos(\beta-\alpha) = 0$, $m_T = 1$ TeV with $\tan\beta = 0.7$ (right) and $\sin\theta_L = V_{Tb} = 10^{-2}$ (left).}
\label{fig:7}
\end{figure}
%%%%%%%%%%%%%%%%%%%%%%%%%%%%%%%%%%%%%
In Fig.~\ref{fig:7} we illustrate the BRs of the $T$ quark as a function of
$\sin\theta_L$ (right panel) and as a function of $\tan\beta$ (left panel). 
We assume  a heavy scalar scenario where  $m_H=m_A=500$ GeV, $H^\pm=600$ GeV 
 and $m_T = 1$ TeV. 
As seen from the right plot for $\tan\beta = 0.7$, $T\to b W^+$ and
$T\to b H^+$ are comparable and could reach 25$\%$ at small mixing
$\sin\theta_L = 0$ instead of 50$\%$ in the SM for $T\to b W^+$. 
However, $T\to t h $ and $tZ$ are slightly smaller compared to SM+VLQ case
 (less than 12\%).  
% (see \cite{Harigaya:2012ir}). In Fig.~\ref{fig:7} (left), we plot again the BRs of the $T$ state  but now as
a function of $\tan\beta $ and with small mixing $\sin\theta_L=10^{-2} $. 
We see that in the limit $\sin\theta_L \to 0$, when the non-standard $T$ decay
modes $T\to tH, tA$ are suppressed due to their coupling, 
which is proportional to $\sin\theta_L$, $T\to bH^+$ is comparable to
  $T\to bW^+$ for $\tan\beta\approx 0.5-0.7$ and 
may offer an alternative discovery mode. We stress here that $b\to s\gamma$ 
constraint is fulfilled for such small $\tan\beta\approx 0.5-0.7$ if  
$m_{H^\pm}=600$ GeV.
%%%%%

\begin{figure}[hptb]%SJDK: uncomment file and vspace
\includegraphics[width=1.05\textwidth]{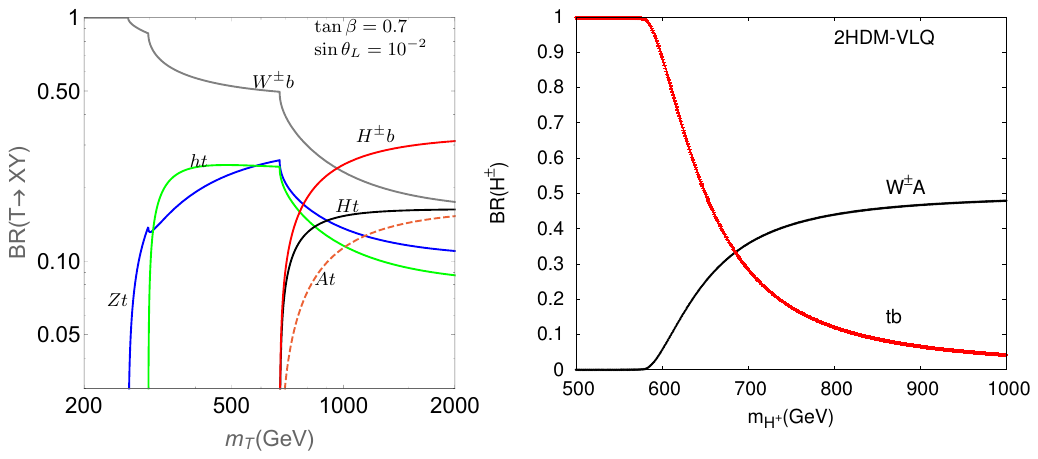}
\vspace{-17cm}
\caption{
Left: BRs of the $T$ as a fucntion of $m_T$ with the same parameter as in Fig.~\ref{fig:7}.
Right: BRs of the $H^\pm$ state in the 2HDM+VLQ as a function   of its mass for $\sin\theta_L=0 $, 
 $y_T = 6$, $m_h = 125.5$ GeV, $m_A = m_H = 500$ GeV, $\cos(\beta-\alpha) = 0$, $\tan\beta = 0.7$ and $m_T = 1000 $ GeV.
(Here, we assume $m_T>m_{H^\pm}$, so that 2HDM-like decays only are included, i.e., the $H^\pm$  state cannot decay into final states with $T$s.) }
\label{fig:HpmBRs}
\end{figure}

Finally, 
in Fig.~\ref{fig:HpmBRs} (left) we illustrate the branching fractions of the heavy
top as a function of $m_T$. As it can be seen, when non-SM decays such as $T\to
bH^+, tH, tA$ are not open, the situation is similar to the SM+VLQ case, as expected. However, when 
$T\to bH^+, tH, tA$ are open, one can see that $T\to bH^+$ compete with 
$T\to bW^+$ and even dominate for high $m_T$. Note that, at large $m_T$,
$T\to tH, tA$ are slightly larger than the SM decays $T\to th, tZ$.
%In Fig.~\ref{fig:HpmBRs} (right) we illustrate branching fractions the
%charged Higgs $m_{H\pm}$.

There exist several LHC analyses searching for (model-independent) pair produced new VLQ states, performed by both ATLAS and CMS.
These place limits in the range 600--800 GeV depending on the actual BR of the $T$ quark in the channels searched for. These do not presently include the $T\to bH^+$ mode. However, 
with $m_T = 1000$ GeV, the $T \bar{T} \to  b\bar{b} H^+ H^-$  mode (followed by $ H^+ H^-\to b\bar b  b\bar{b}
W^+W^-$ decays) can lead to `$W^+ W^-$ plus $6b$-jets'  as a sizeable and (very distinctive) signature. Even with small $\sin\theta_L $, an interesting possibility would be $T \bar{T} \to  b\bar b W^{+}H^{-}$ decays, with $H^-\to  b\bar b W^-$,
producing an equally distinctive `$W^+ W^-$ plus $4b$-jets' signal. While  
also the `$W^{+*}W^{-*}W^+W^-$ plus $4b$-jets' case is a potentially interesting channel, 
stemming from $T\bar T\to t\bar t hh\to b\bar bW^+W^-hh$ with  $hh\to b\bar b W^{+*}W^{-*}$ decays,
given that $T\to t h $ is probably very difficult to detect  (as intimated already, see \cite{Harigaya:2012ir}), the alternative
of accessing  the new VLQ state via $T \to b H^+ $ decays becomes a very intriguing one. In fact, this is the ideal  channel to characterize our model, as even neutral $A,H$ decays are, so-to-say, `degenerate' (i.e., they can have the same decay patterns) with those from the SM-like Higgs state, $h$. Needless to say, to establish this signature would represent circumstantial evidence of a 2HDM+VLQ structure. Typical $H^\pm$ decay patterns can be found in Fig.~\ref{fig:HpmBRs} (right), showing that $b\bar b W^\pm$ decays of a charged Higgs bosons (via $tb, W^\pm A$) are indeed the dominant ones for large $m_{H^\pm}$ values.

%%%%%%%%%%%%%%%%%%%%%%%%%%%%%%%%%%%%%%
%%%%%%%%%%%%%%%%%%%%%%%%%%%%%%%%%%%%%%
%%%%%%%%%%%%%%%%%%%%%%%%%%%%%%%%%%%%%
%%%%%%%%%%%%%%%%%%%%%%%%%%%%%%%%%%%%%%%%
%
\section{Conclusions}
\label{sec:VI}

In this paper, we have extended the ordinary 2HDM-II by a singlet 
heavy VLQ  with the same EM
charge as the top quark. In the (near) decoupling limit of the 2HDM-II, 
one neutral CP-even Higgs state, $h$,  can mimic the
SM-like Higgs boson seen at the LHC Run 1 at $\approx 125$ GeV  while the other two
neutral Higgs states, $H$ and $A$, can be constrained as the model has to accommodate the 
data observed by ATLAS and CMS at the LHC Run 2 in both the low and 
high mass region, which are to date consistent with SM expectations.  We have then proceeded to a phenomenological comparison between the 2HDM+VLQ scenario and the ordinary 2HDM case, limitedly to the LHC environment.

We have  illustrated that a different decay pattern emerges in the 2HDM+VLQ with respect to the standard 2HDM when $\gamma\gamma$ and $Z\gamma$ samples are compared to each other. Then, we have shown that, at the same time, an enhancement of  $\sigma(pp\to t\bar{t}h)$  by a factor up to 2 can occur, which would explain an  increased value of  such a cross-section at the LHC, i.e., in the direction of a possible enhancement seen in  Run 2 data.
All this can occur for VLQ masses of order 600--800 GeV, so that we have finally
 highlighted that non-SM-like decays of this VLQ state, particularly via $H^\pm$ channels, 
would be  evidence of a 2HDM sector. 

In fact, by combining these  instances,
a peculiar `smoking-gun' situation may emerge in the 2HDM+VLQ scenario discussed here, where
one has $\gamma\gamma$ rates at large invariant masses essentially compatible with the SM background,
yet a depletion (with respect to the 2HDM yield) can be seen in the $Z\gamma$ sample,
with or without an enhancement of the $t\bar t h$ cross-section. One could indeed disentangle this as being due to this particular BSM structure by finally revealing a variety of $T\to bH^+$ decays emerging from QCD induced $T\bar T$ production.

Remarkably, all such phenomenology can be obtained for parameter space configurations compliant
with current theoretical and experimental constraints, as we have  scrupulously assessed using up-to-date tools, yet ameanable to prompt phenomenological investigation in the upcoming years at the LHC, during Run 2 and 3.

\appendix

%==============================
\section{Yukawa couplings in the 2HDM+VLQ}
%================================
%\footnote {We are interested in the couplings of scalar
%Higgs bosons $h,H$ and $A$ to the
%SM fermions of the third generation and \textcolor{red}{the gauge bosons}
%parametrised in terms of the scale factors}
These are as follows:
\begin{eqnarray}
\kappa^h_{t\bar{t}} &=& c_L c_R (y^h_{t}) - s_R c_L \left(c_{\beta\alpha}\frac{y_T}{y_t} + s_{\beta\alpha}\frac{\xi_T}{y_t}\right),\\
\kappa^h_{T\bar{T}} &=& s_L s_R (y^h_{t}) + s_L c_R \left(c_{\beta\alpha}\frac{y_T}{y_t} + s_{\beta\alpha}\frac{\xi_T}{y_t}\right), \\
%Note in the "Verification-eq-21--29.nb" file, there is a mistake
%for this eqn, it should be tTh; not tTH as written -Simon
%
\kappa^h_{t\bar{T}} &=& c_R s_L (y^h_{t}) - s_R s_L \left(c_{\beta\alpha}\frac{y_T}{y_t} + s_{\beta\alpha}\frac{\xi_T}{y_t}\right),\\
\kappa^H_{t\bar{t}} &=& c_L c_R (y^H_{t}) + s_R c_L \left(s_{\beta\alpha}\frac{y_T}{y_t} - c_{\beta\alpha}\frac{\xi_T}{y_t}\right),\\
\kappa^H_{T\bar{T}} &=& s_L s_R (y^H_{t}) - s_L c_R \left(s_{\beta\alpha}\frac{y_T}{y_t} - c_{\beta\alpha}\frac{\xi_T}{y_t}\right),\\
\kappa^H_{t\bar{T}} &=& c_R s_L (y^H_{t}) + s_R s_L \left(s_{\beta\alpha}\frac{y_T}{y_t} - c_{\beta\alpha}\frac{\xi_T}{y_t}\right),\\
\kappa^A_{t\bar{t}} &=& i\left(c_L c_R (y^A_{t}) + s_R c_L \frac{y_T}{y_t}\right),\\
\kappa^A_{T\bar{T}} &=& i \left(s_L s_R (y^A_{t}) - s_L c_R \frac{y_T}{y_t}\right),\\
\kappa^A_{t\bar{T}} &=& i \left(s_L c_R (y^A_{t}) + s_R s_L \frac{y_T}{y_t}\right).\\
\nonumber
\end{eqnarray}
%=======================================
%=======================================
The above reduced Higgs couplings $\kappa^{\phi}_{ij}$ are expressed in terms
 of the normalized $y^{\phi}_{t}$ ones given by
\begin{eqnarray}
y^{h}_{t} &=& \sin(\beta -\alpha) + \cot\beta\cos(\beta -\alpha), \nonumber\\
y^{H}_{t} &=& \cos(\beta -\alpha) - \cot\beta\sin(\beta -\alpha), \nonumber\\
y^{A}_{t} &=& \cot\beta.
%\nonumber
\end{eqnarray}
It is easy to check that,
in the case of zero mixing $s_L=0$, the $\{h,H,A\} t\bar{t}$ couplings reduce to the
2HDM ones while $\{h,H,A\} t\bar{T}$ and $\{h,H,A\} T\bar{T}$ all vanish.
%==============================
\section{Form factors for $b\to s \gamma$}
%==============================
These were used as follows:
 \begin{align}
 f_{1\gamma}(x) & = \frac{x}{72} \left[ \frac{8 x^2 + 5 x -7}{(1-x)^3} - \frac{6 x (2-3x)}{(1-x)^4} \ln(x)\right]\,, \nonumber \\
f_{1G}(x) & = \frac{x}{24} \left[ \frac{ x^2 - 5 x -2}{(1-x)^3} - \frac{6 x }{(1-x)^4} \ln(x)\right]\,, \nonumber \\
 f_{2\gamma}(x) & = \frac{x}{12}  \left[ \frac{3-5x}{(1-x)^2} + \frac{2(2-3x)}{(1-x)^3 }\ln(x)  \right]\,, \nonumber \\
f_{2G}(x) & = \frac{x}{4}  \left[ \frac{3-x}{(1-x)^2} + \frac{2}{(1-x)^3 } \ln(x)\right]\,.
 \end{align}

%==============================
\section{Partial widths of the VLQ}
%================================
In this appendix we give the analytic expressions of the partial widths of the VLQ into vector and Higgs bosons,  
$T\to q V$ and 
$T\to q\phi $, as  
\begin{eqnarray}
\Gamma(T\to q V) &=& \frac{g^2}{32\pi} \frac{\beta}{m^2_T}\left((g^2_L + g^2_R) \left( \frac{m^2_T + m^2_q}{2} + 
\frac{(m^2_T - m^2_q)^2}{2m^2_V}  - m^2_V\right) - 6g_L g_R m_T m_q\right), \\ 
\Gamma(T\to q \phi ) &=& \frac{\beta}{8\pi m^2_T} \left( (g^2_L + g^2_R)\frac{m^2_T + m^2_q - m^2_{\phi}}{2}  + g_L g_R m_T m_q \right),
\end{eqnarray}
with 
\begin{eqnarray}
\beta =  \frac{\left((m^2_T -(m_q + m_X)^2) (m^2_T -(m_q - m_X)^2)\right)^p}{2m_T} \quad \quad (X = V, \phi),
\end{eqnarray}
where $m_V$($m_\phi$) and $m_q$ are the masses of the gauge(Higgs) bosons and the SM quark, respectively. We denote as $g_L$ and $g_R$ the left- and right-handed components of the SM quark $q$. Finally, $p = 1/2$ for $\phi = h, H$ and $p = 3/2$ for $\phi = A$.

\newpage
\begin{acknowledgments}
The authors are supported by the grant H2020-MSCA-RISE-2014 no. 645722
(NonMinimalHiggs).
This work is also supported by the Moroccan Ministry of Higher
Education and Scientific Research MESRSFC and  CNRST: Project PPR/2015/6.
SM is supported in part through the NExT Institute.
RB and CSU thank SM for hospitality in Southampton when part of this research was
carried out. AA and RB would also 
like to acknowledge the hospitality of the National Center for Theoretical
Sciences (NCTS), Physics Division, in Taiwan.

\vspace{-0.5cm}
\end{acknowledgments}

%%%%%%%%%%%%%%%%%%%%%%%%%%%%%%%%%%%%%%%%

%%%%%%%%%%%%%%%%%%%%%%%%%%%%%%%%%%%%%%%%
\end{document}